\documentclass[useAMS,usenatbib]{mn2e}
\usepackage[dvips]{graphicx}
\usepackage{url}
\usepackage{color}
\newcommand{\bz}{$\langle B_z \rangle$}
\newcommand{\nz}{$\langle N_z \rangle$}
\newcommand{\vsini}{$v \sin i$}
\newcommand{\kms}{km\,s$^{-1}$}

\newcommand{\mdot}{$\dot{M}$}

\newcommand{\msun}{$M_\odot$}

\newcommand{\teff}{$T_{\rm eff}$}
\newcommand{\ra}{$R_{\rm A}$}
\newcommand{\rk}{$R_{\rm K}$}

\title[NU Ori: a hierarchical SB3 with a magnetic B star]{NU Ori: a hierarchical triple system with a strongly magnetic B-type star}
\author[M. Shultz]
{M.\ Shultz$^{1,2}$\thanks{E-mail: matt.shultz@gmail.com},
J.-B.\ Le Bouquin$^{3}$,
Th.\ Rivinius$^{4}$,
G.\ A.\ Wade$^{5}$,
O.\ Kochukhov$^{1}$,
\newauthor{E.\ Alecian$^{3,6,7}$, V.\ Petit$^{8}$, O.\ Pfuhl$^{9}$, M.\ Karl$^{9}$, F.\ Gao$^{9}$, R.\ Grellmann$^{10}$, C.-C.\ Lin$^{11}$,}
\newauthor{P.\ Garcia$^{3,12}$, S.\ Lacour$^{6}$, and the MiMeS and BinaMIcS Collaborations} \\
$^1$Department of Physics and Astronomy, Uppsala University, Box 516, Uppsala 75120, Sweden \\
$^2$Annie Jump Cannon Fellow, Department of Physics and Astronomy, University of Delaware, 217 Sharp Lab, Newark, Delaware, 19716, USA\\
$^3$Universit\'e Grenoble Alpes, IPAG, F-38000 Grenoble, France\\
$^4$ESO - European Organisation for Astronomical Research in the Southern Hemisphere, Casilla 19001, Santiago 19, Chile\\
$^5$Department of Physics and Space Science, Royal Military College of Canada, Kingston, Ontario K7K 7B4, Canada\\
$^6$CNRS, IPAG, F-38000 Grenoble, France\\
$^7$LESIA, Observatoire de Paris, CNRS UMR 8109, UPMC, Universit\'e Paris Diderot, 5 place Jules Janssen, 92190, Meudon, France\\
$^8$Department of Physics and Astronomy, University of Delaware, 217 Sharp Lab, Newark, Delaware, 19716, USA\\
$^9$Max Planck Institute for Extraterrestrial Physics, Giessenbachstrasse 1, 85748 Garching, Bayern, Germany\\
$^{10}$Physikalisches Institut der Universit\"at zu K\"oln, Z\"ulpicher Str.\ 77, 50397 K\"oln, Germany\\
$^{11}$Max Planck Institute for Astronomy, K\"onigstuhl 17, 69117 Heidelberg, Germany\\
$^{12}$Centro de Astrof\'isica e Gravita\c{c}\~{a}o, IST, Universidade de Lisboa, P-1049-001 Lisboa, Portugal\\
Faculdad de Engenharia, Universidade de Porto, rua Dr. Roberto Frias, 4200-465 Porto, Portugal\\
}
\begin{document}

\date{}

\pagerange{\pageref{firstpage}--\pageref{lastpage}} \pubyear{2002}

\maketitle

\label{firstpage}

\begin{abstract}
NU Ori is a massive spectroscopic and visual binary in the Orion Nebula Cluster, with 4 components: Aa, Ab, B, and C. The B0.5 primary (Aa) is one of the most massive B-type stars reported to host a magnetic field. We report the detection of a spectroscopic contribution from the C component in high-resolution ESPaDOnS spectra, which is also detected in a Very Large Telescope Interferometer (VLTI) dataset. Radial velocity (RV) measurements of the inner binary (designated Aab) yield an orbital period of 14.3027(7)~d. The orbit of the third component (designated C) was constrained using both RVs and interferometry. We find C to be on a mildly eccentric 476(1)~d orbit. Thanks to spectral disentangling of mean line profiles obtained via least-squares deconvolution we show that the Zeeman Stokes $V$ signature is clearly associated with C, rather than Aa as previously assumed. The physical parameters of the stars were constrained using both orbital and evolutionary models, yielding $M_{\rm Aa} = 14.9 \pm 0.5$~\msun, $M_{\rm Ab} = 3.9 \pm 0.7$~\msun, and $M_{\rm C} = 7.8 \pm 0.7$~\msun. The rotational period obtained from longitudinal magnetic field \bz~measurements is $P_{\rm rot} = 1.09468(7)$~d, consistent with previous results. Modeling of \bz~indicates a surface dipole magnetic field strength of $\sim 8$~kG. NU Ori C has a magnetic field strength, rotational velocity, and luminosity similar to many other stars exhibiting magnetospheric H$\alpha$ emission, and we find marginal evidence of emission at the expected level ($\sim$1\% of the continuum). 
\end{abstract}


\section{Introduction}

NU Ori (HD 37061, Brun 747) is the central ionizing B0.5 V star of the M43 H~{\sc ii} region of the Orion Nebula Cluster. The primary is the second hottest B-type star in which a magnetic field has been reported \citep{2008MNRAS.387L..23P}, after $\tau$ Sco \citep{2006MNRAS.370..629D}. The system is both a spectroscopic binary, with an orbital period of between 8 and 19 d \citep{1991ApJ...367..155A,1991ApJS...75..965M}, and a visual binary with two companions, one (designated B) detected by high-contrast adaptive optics \citep{2006A&A...458..461K}, and a second (designated C) via interferometry \citep{2013A&A...550A..82G}. 

As a very young, hot star with a magnetic field and a closely orbiting companion, NU Ori is of interest to investigations of the origin of fossil magnetic fields. The rarity of close magnetic binaries \citep[less than 2\% of close hot binaries contain a magnetic star;][]{2015IAUS..307..330A} has been invoked as supporting evidence by two fossil field formation hypotheses: concentration of primordial magnetic flux within the star-forming environment \citep[in which highly magnetized pre-stellar cores inhibit cloud fragmentation;][]{2011ApJ...742L...9C}, or dynamos powered by binary mergers \citep{2016MNRAS.457.2355S}. The system is of potential relevance to investigations of the possible contribution of magnetic fields to spin-orbit interactions. Finally, as one of the hottest and most rapidly rotating magnetic early B-type stars known \citep{petit2013,2018MNRAS.475.5144S}, it is interesting from the point of view of magnetic wind confinement \citep[e.g.][]{bm1997,ud2002,town2005c}. In particular, despite its powerful wind and rapid rotation, it shows no sign of magnetospheric emission, thus presenting a potential challege to theories of magnetosphere formation in early-type stars.

While an orbital period was published by \cite{1991ApJ...367..155A}, at that time NU Ori was considered to be an SB1. The spectral contribution of the secondary was first reported by \cite{2008MNRAS.387L..23P}, however, an orbital model making use of the secondary's radial velocities has yet to be published. Since the Zeeman Stokes $V$ signatures are much wider than the secondary's narrow line profiles, \citeauthor{2008MNRAS.387L..23P} inferred the broad-lined primary to be the magnetic star. While the surface dipolar magnetic field strength was constrained via Bayesian analysis of the circular polarization profile by \cite{2008MNRAS.387L..23P}, the rotational period of the magnetic star was not known at that time, making a precise magnetic model difficult to determine. Using an expanded spectropolarimetric dataset, a rotational period of $\sim1.1$~d was determined by \cite{2018MNRAS.475.5144S} using longitudinal magnetic field measurements. 

\cite{2017MNRAS.471.2286S} noted that the primary star of the NU Ori system has stellar and magnetospheric parameters very similar to those of the magnetic $\beta$ Cep variable $\xi^1$ CMa, and yet, unlike $\xi^1$ CMa, shows no sign of magnetospheric emission. The two principal differences between the stars, from a magnetospheric standpoint, are 1) NU Ori's very rapid rotation (\vsini~$\sim 200$~\kms, $P_{\rm rot} \sim 1.1$~d) as compared to $\xi^1$ CMa's extremely slow rotation ($P_{\rm rot} \sim 30$~yr), and 2) NU Ori's status as a close binary. At least one star, HD 156324, is known to have a magnetosphere strongly disrupted by orbital dynamics \citep{2018MNRAS.475..839S}, making it natural to wonder if the failure to detect emission around this star might also be due to the presence of a nearby orbiting companion. This motivates a closer examination of NU Ori's rotational, magnetic, and orbital parameters. 

The paper is organized as follows. Observations are described in \S~\ref{sec:obs}. Radial velocity measurements, orbital period determination, and orbital modeling are presented in \S~\ref{sec:multiplicity}. \S~\ref{sec:mag} presents the magnetometry. In \S~\ref{sec:discussion} orbital parameters are used to constrain stellar parameters, which are in turn used with the magnetic and rotational parameters to investigate the star's magnetic, rotational, and magnetospheric properties. The conclusions are summarized in \S~\ref{sec:conclusions}.

\section{Observations}\label{sec:obs}

\subsection{Spectropolarimetry}

ESPaDOnS is a fibre-fed echelle spectropolarimeter mounted at the Canada-France-Hawaii Telescope (CFHT). It has a spectral resolution $\lambda/\Delta\lambda \sim 65,000$, and a spectral range from 370 to 1050 nm over 40 spectral orders. Each circular polarization observation consists of 4 polarimetric sub-exposures, between which the orientation of the instrument's Fresnel rhombs are changed, yielding 4 intensity (Stokes $I$) spectra, 1 circularly polarized (Stokes $V$) spectrum, and 2 null polarization ($N$) spectra \citep[defined by][]{d1997}. \cite{2016MNRAS.456....2W} described the reduction and analysis of ESPaDOnS data in detail together with the observing strategy of the Magnetism in Massive Stars (MiMeS) Large Programs (LPs). 

The first 6 observations were acquired in 2006 and 2007, and were reported by \cite{2008MNRAS.387L..23P}\footnote{Program codes CFHT 05C24 and 07AC10.}. Fourteen additional Stokes $V$ observations were acquired between 01/2010 and 02/2012 by the MiMeS LP, and a further 10 observations between 11/2012 and 12/2012 by a P.I. program\footnote{Program Code CFHT 14AC010}. Sub-exposure times varied between 700 and 800 s; the full exposure time is 4$\times$ this duration. The median peak signal-to-noise (S/N) per spectral pixel is 993. Two observations with a S/N~$\sim 100$ were discarded from the magnetic analysis (although these could still be used to obtain radial velocity measurements for the primary). The observation log is provided in Table \ref{rvtab}.

\begin{table}
\centering
\caption{Observation log and RV measurements. Estimated RV uncertainties are 6~\kms~for Aa, 2~\kms~for Ab, and 9~\kms~for C.}
\resizebox{8.5 cm}{!}{
\label{rvtab}
\begin{tabular}{l l r r r r r}
\hline
\hline
\\
Cal. & HJD        & $t_{\rm exp}$ & S/N & RV (Aa) & RV (Ab) & RV (C) \\
Date & -2450000   & (s)           &    & (\kms)  & (\kms)  & (\kms) \\
\hline
2006-01-12 & 3747.87907 & 4$\times$800 &  942 &  39 &   53 & -11 \\
2006-01-12 & 3747.92079 & 4$\times$800 &  923 &  39 &   50 & -15 \\
2006-01-12 & 3747.96263 & 4$\times$800 &  957 &  41 &   47 & -17 \\
2007-03-08 & 4167.72958 & 4$\times$800 & 1066 &  79 &  -87 &  -7 \\
2007-03-08 & 4167.76959 & 4$\times$800 & 1081 &  79 &  -84 &  -2 \\
2007-03-08 & 4167.81034 & 4$\times$800 &  993 &  80 &  -83 &  -6 \\
2010-01-26 & 5222.77964 & 4$\times$800 & 1142 &  69 &  -88 & -21 \\
2010-01-26 & 5222.81932 & 4$\times$800 & 1150 &  70 &  -88 & -19 \\
2010-02-01 & 5228.75094 & 4$\times$800 &  963 &  19 &   93 &   4 \\
2010-02-01 & 5228.79260 & 4$\times$800 &  976 &  19 &   97 &  -4 \\
2010-02-02 & 5229.79356 & 4$\times$800 & 1130 &   1 &  156 &   4 \\
2010-02-02 & 5229.83319 & 4$\times$800 & 1171 &   1 &  158 &   4 \\
2011-11-05 & 5871.10842 & 4$\times$740 &  261 &  15 &  -30 &  79 \\
2011-11-06 & 5872.08868 & 4$\times$740 & 1017 &  -1 &   39 &  71 \\
2011-11-06 & 5872.15256 & 4$\times$740 & 1177 &  -4 &   43 &  71 \\
2011-11-08 & 5874.06553 & 4$\times$740 &  111 & -33 &  -- & -- \\
2011-11-09 & 5875.03292 & 4$\times$740 & 1147 & -44 &  176 &  75 \\
2012-01-04 & 5930.71891 & 4$\times$740 &  967 & -32 &  124 &  75 \\
2012-01-17 & 5943.90765 & 4$\times$740 & 1250 & -12 &   59 &  80 \\
2012-02-04 & 5961.75237 & 4$\times$800 & 1367 & -45 &  180 &  56 \\
2012-11-26 & 6257.90822 & 4$\times$700 &   83 &   5 &  -- & -- \\
2012-11-26 & 6257.94322 & 4$\times$700 &  306 &  21 &   33 &  39 \\
2012-12-05 & 6266.86057 & 4$\times$700 &  226 &  59 &  -99 &  30 \\
2012-12-05 & 6266.90271 & 4$\times$700 &  605 &  57 & -103 &  35 \\
2012-12-05 & 6266.93881 & 4$\times$700 &  366 &  57 & -103 &  50 \\
2012-12-22 & 6283.97259 & 4$\times$700 & 1182 &  62 & -139 &  52 \\
2012-12-25 & 6286.92096 & 4$\times$700 &  898 &   5 &   57 &  46 \\
2012-12-25 & 6286.95840 & 4$\times$700 &  451 &   6 &   57 &  44 \\
2012-12-25 & 6286.99299 & 4$\times$700 & 1041 &   3 &   59 &  45 \\
2012-12-28 & 6289.78387 & 4$\times$700 & 1087 & -33 &  191 &  48 \\
\hline
\hline
\end{tabular}
}
\end{table}

\subsection{Interferometry}\label{subsec:interf_obs}

NU Ori was observed with the PIONIER\footnote{Program codes 60.A-0209(A), 092.C-0542(A), 094.C-0175(A), 094.C-0397(A), and 096.D-0518(A).} \citep{2011A&A...535A..67L} and GRAVITY\footnote{Program code 0100.C-0597(A).} \citep{2017A&A...602A..94G} instruments from the Very Large Telescope Interferometer \citep{2014SPIE.9146E..0JM}. The PIONIER data were reduced and calibrated with the {\sc pndrs} package from the JMMC\footnote{http://www.jmmc.fr/pndrs}. The GRAVITY data were reduced and calibrated with the ESO package. All observations spatially resolved the C component of the system; the B component was not detected. The GRAVITY data were also reported by \cite{gravity2018} in their survey of multiplicity in the Orion Nebula Cluster. 

\begin{table}\centering
\caption{Interferometric observations of the AC pair. The separation and position angle are the position of the secondary (faintest in H band) with respect to the primary (brightest in H band), measured east (90~deg) from north (0~deg). The columns e$_{\rm max}$ and e$_{\rm min}$ are the FWHM of the major and minor axes of the astrometric error ellipse. The column P.A.~e$_{\rm max}$ is the position angle of the major axis.}  
\resizebox{8.5 cm}{!}{
\label{tab:interfero}
 \begin{tabular}{cccccccccccc}
   \hline\hline\noalign{\smallskip}
   Instrument & MJD & Obs. & Sep.  & P.A.  & e$_{\rm max}$ & e$_{\rm min}$ & P.A.~e$_{\rm max}$ \\
              &     & No.  & (mas) & (deg) & (mas) & (mas) & (deg) \\ \hline
   \noalign{\smallskip}\hline\noalign{\smallskip}
   PIONIER & 56601.198 & 1 & 8.75 &  -85.27 &  0.43 &  0.35 &   5 \\
   PIONIER & 56991.324 & 2 & 3.60 & -128.41 &  0.59 &  0.33 &  69 \\
   PIONIER & 56994.236 & 3 & 4.04 & -124.45 &  0.20 &  0.07 & 152 \\
   PIONIER & 57007.196 & 4 & 5.37 & -112.31 &  0.26 &  0.10 & 157 \\
   PIONIER & 57382.175 & 5 & 7.74 &  +93.01 &  0.55 &  0.40 & 102 \\
   PIONIER & 57386.128 & 6 & 7.51 &  +93.49 &  0.74 &  0.34 &   5 \\
   GRAVITY & 58039.3   & 7 & 8.543 & -82.94 &  0.02 &  0.02 & 5 \\
   GRAVITY & 58039.4   & 8 & 8.549 & -82.84 &  0.02 &  0.02 & 5 \\
   GRAVITY & 58128.1   & 9 &  4.39 & -37.86 &  0.02 &  0.02 & 5 \\
   GRAVITY & 58128.1   & 10 &  4.40 & -37.76 &  0.02 &  0.02 & 5 \\
   \noalign{\smallskip}\hline
 \end{tabular}
}
\end{table}

\section{Multiplicity}\label{sec:multiplicity}

\subsection{Radial velocities}\label{sec:rv}

   \begin{figure}
   \centering
   \includegraphics[width=\hsize]{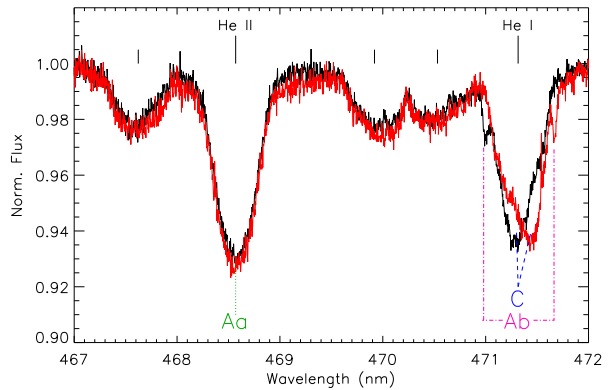}
      \caption[]{Spectra in the vicinity of the He~{\sc ii} 468.6 nm line at maximum RV separation of the Aab pair. The black spectrum was acquired on 2012-12-22 and the red spectrum on 2012-12-28. Unlabelled lines indicate O~{\sc ii} lines. The spectra have been shifted to the rest velocity of the Aa component. There is no line profile variability in He~{\sc ii} 468.6 nm. The adjacent He~{\sc i} 471.3 nm line shows a contribution from the Ab component, however it also shows variation that cannot be attributed to either Aa or Ab. This is due to the contribution of a third star, designated C.}
         \label{NUOri_HeII4686_HeI4713_spec}
   \end{figure}

   \begin{figure}
   \centering
   \includegraphics[width=\hsize]{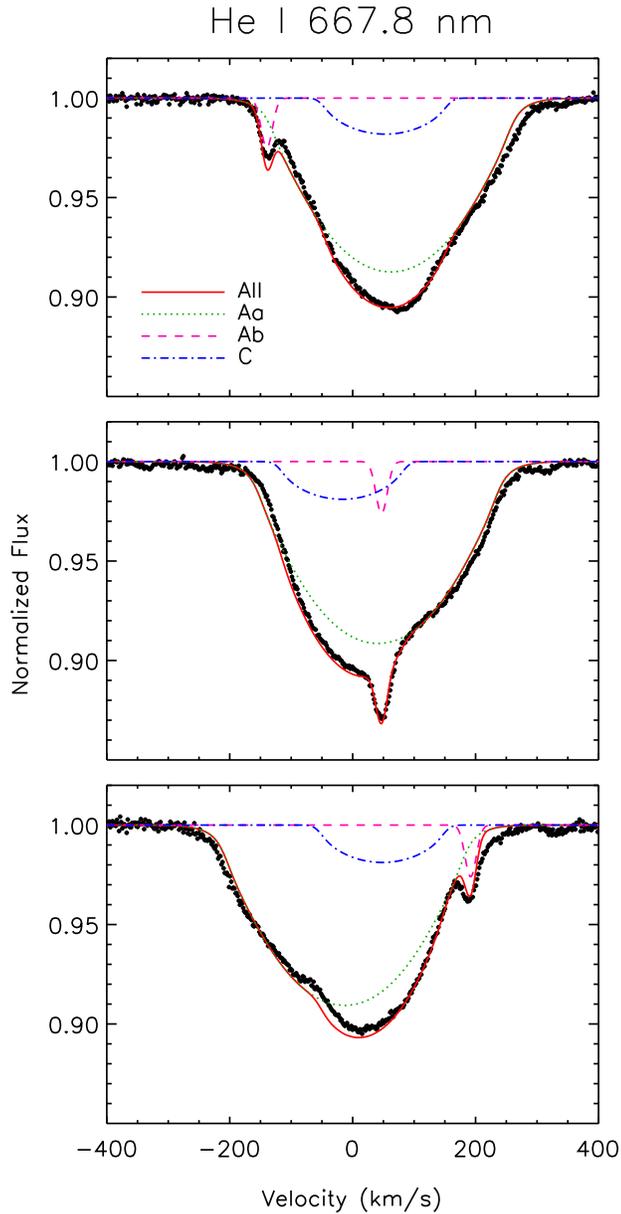}
      \caption[]{He~{\sc i} 667.8 nm line with a 3-component fit at 3 representative RV separations of the Aab components.}
         \label{NUOri_HeI6678_3starfit}
   \end{figure}

While NU Ori's spectrum is dominated by the hot primary star, it is a known SB2 \citep{2008MNRAS.387L..23P}, with contributions from a sharp-lined secondary. However, its orbital period is not yet well-constrained. We respectively designate the primary and secondary as Aa and Ab, following the nomenclature adopted by \cite{2013A&A...550A..82G}. Radial velocities (RVs) were measured for the broad-lined primary using the He~{\sc ii} 468.6 nm line, which shows no contribution from the secondary. Measurements were obtained by fitting synthetic line profiles convolved with rotational and turbulent broadening using the binary line profile fitting routine described by \cite{2017MNRAS.465.2432G}, although for the He~{\sc ii} 468.6 nm line only a single stellar component was utilized. These RVs are provided in Table \ref{rvtab}. In the process, we also obtained \vsini~and macroturbulent $v_{\rm mac}$ measurements. 

Representative spectra in the vicinity of He~{\sc ii} 468.6 nm are shown in Fig.\ \ref{NUOri_HeII4686_HeI4713_spec}, where the spectra have been shifted to the rest velocity of the Aa component. He~{\sc ii} 468.6 nm shows no intrinsic line profile variability. The same is true for nearby O~{\sc ii} lines, which are also expected to be formed solely in the Aa component's photosphere since it should be much hotter than Ab (Aa contributes about 20$\times$ as much flux as Ab and has an \teff~of $\sim 30$ kK, \citealt{2011AA...530A..57S}; given the flux ratio, Ab should have an \teff~of $\sim 15$~kK, in which O~{\sc ii} lines should not form). In contrast, the nearby He~{\sc i} 471.3 nm line shows a complex pattern of line profile variations. This can be partly attributed to the narrow-lined secondary Ab component, however, this star's contribution is not able to fully explain the variability.

Close inspection of the He~{\sc i} lines demonstrated that the line profile variations can be explained by the presence of a third component, which itself has a variable RV. Fig.\ \ref{NUOri_HeI6678_3starfit} shows 3 representative observations of the He~{\sc i} 667.8 nm line, which was selected for analysis as it is both strong and isolated. The B component is approximately 0.47'' from Aa, and the C component is separated by about 0.015'', thus both would have been within the 1.8'' ESPaDOnS aperture. However, the B component is extremely faint compared to Aa \citep[$\Delta K = 3.23$~mag;][]{2006A&A...458..461K}, and so unlikely to contribute much flux. Furthermore, at a distance of $\sim$400 pc, the B component's projected separation is about 190 AU, which would indicate an orbital period of centuries; RV variation on a timescale of a few years is thus unlikely. We therefore attribute the third spectroscopic component to NU Ori C. 

The line can be fit using a 3-star model: Aa with \vsini~$=190 \pm 10$~\kms~and $v_{\rm mac} = 100 \pm 20$~\kms (see the following paragraph), Ab with \vsini~$=10 \pm 5$~\kms~and $v_{\rm mac} = 5 \pm 5$~\kms, and C with \vsini~$=100 \pm 10$~\kms~and $v_{\rm mac} = 5 \pm 5$~kms. The fractional contributions of the best-fit synthetic line profiles to the total equivalent width are 88\% for Aa, 2\% for Ab, and 10\% for C. RV measurements of Ab and C were obtained from 3-component fits to He~{\sc i} 667.8 nm, and are provided in Table \ref{rvtab} (except for two observations for which the S/N was insufficient to clearly distinguish these lines). RV uncertainties were determined statistically, by fitting the same observations using different initial guesses for the full width at half-maximum of the individual components, and taking the standard deviation of the results. Uncertainties are estimated at 6~\kms~for Aa, 2~\kms~for Ab, and 9~\kms~for C.

The macroturbulence determined for Aa is on the high end of the observed distribution for early B-type stars \citep{2014A&A...562A.135S,2017A&A...597A..22S}. This is likely an artifact of Stark broadening, which was not taken account of in the model. \cite{2014A&A...562A.135S} found compatible values of both \vsini~and $v_{\rm mac}$ using the Si~{\sc iii} 455.3 nm line, however they used a 1-star model and this line exhibits clear variability indicative of a significant contribution from the C component. Using the Si~{\sc iii} line, but disentangling the line profiles of the 3 components as described below in \S~\ref{sec:mag}, we find the same \vsini~but $v_{\rm mac} = 20 \pm 10$~\kms. 

\subsection{Period analysis}\label{subsec:orb_periods}

   \begin{figure}
   \centering
   \includegraphics[width=\hsize]{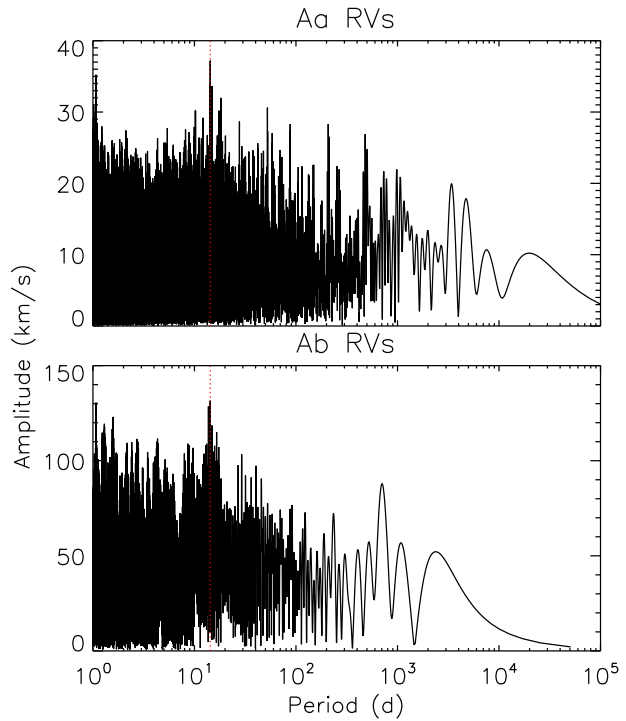}
      \caption[]{Periodograms for Aa and Ab RVs. The dotted red line indicates the maximum amplitude period.}
         \label{Aab_periods}
   \end{figure}

   \begin{figure}
   \centering
   \includegraphics[width=\hsize]{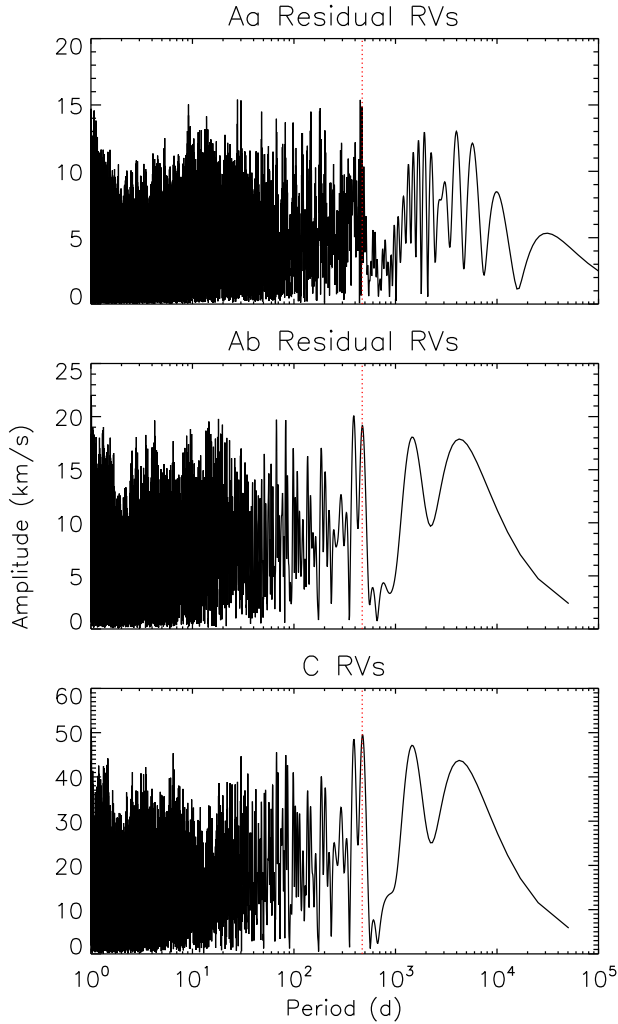}
      \caption[]{Periodograms for Aa and Ab residual RVs, and C RVs. The dotted red line indicates the most significant period from the Aa and C measurements.}
         \label{AB_periods}
   \end{figure}

   \begin{figure}
   \centering
   \includegraphics[width=\hsize]{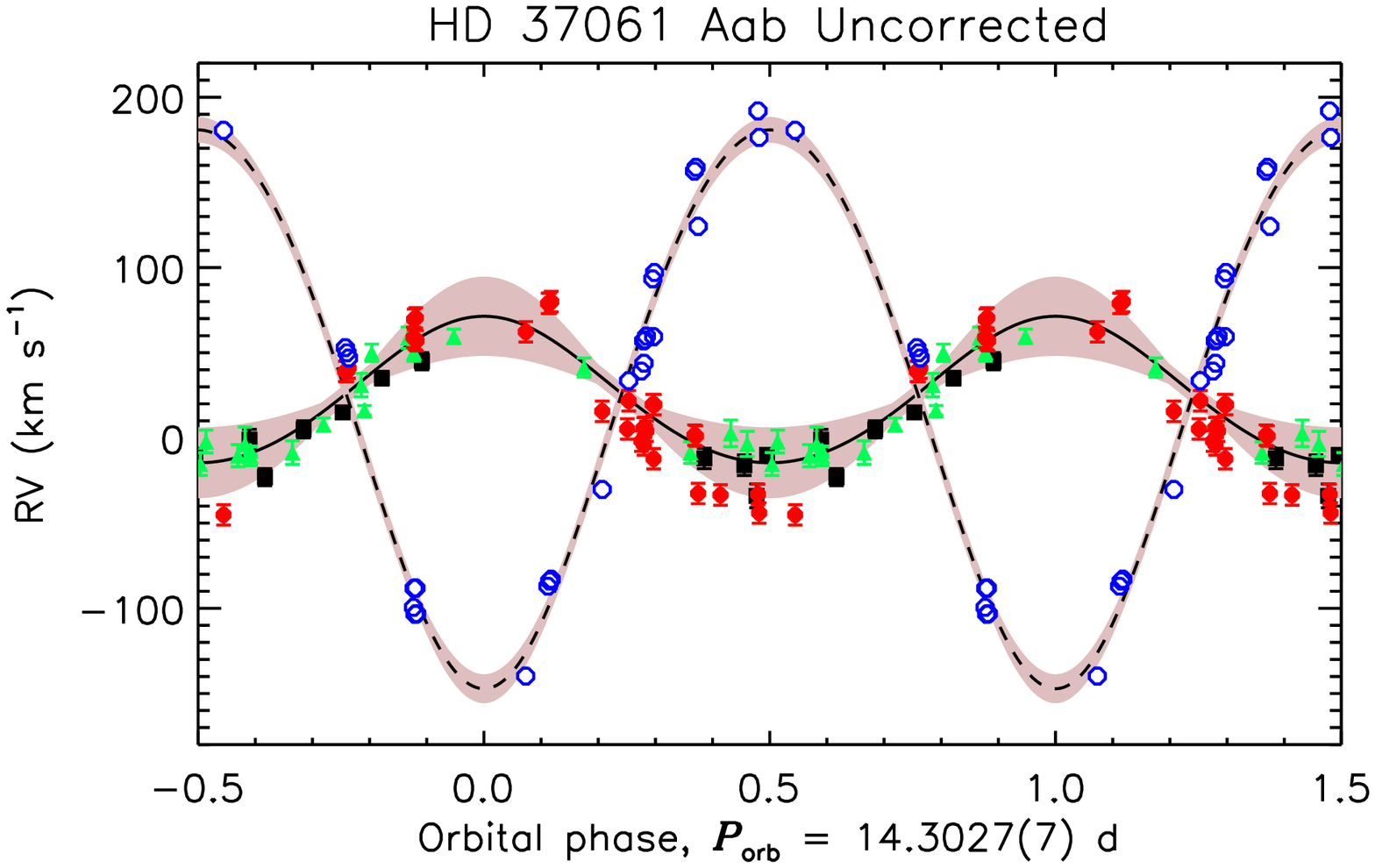}
   \includegraphics[width=\hsize]{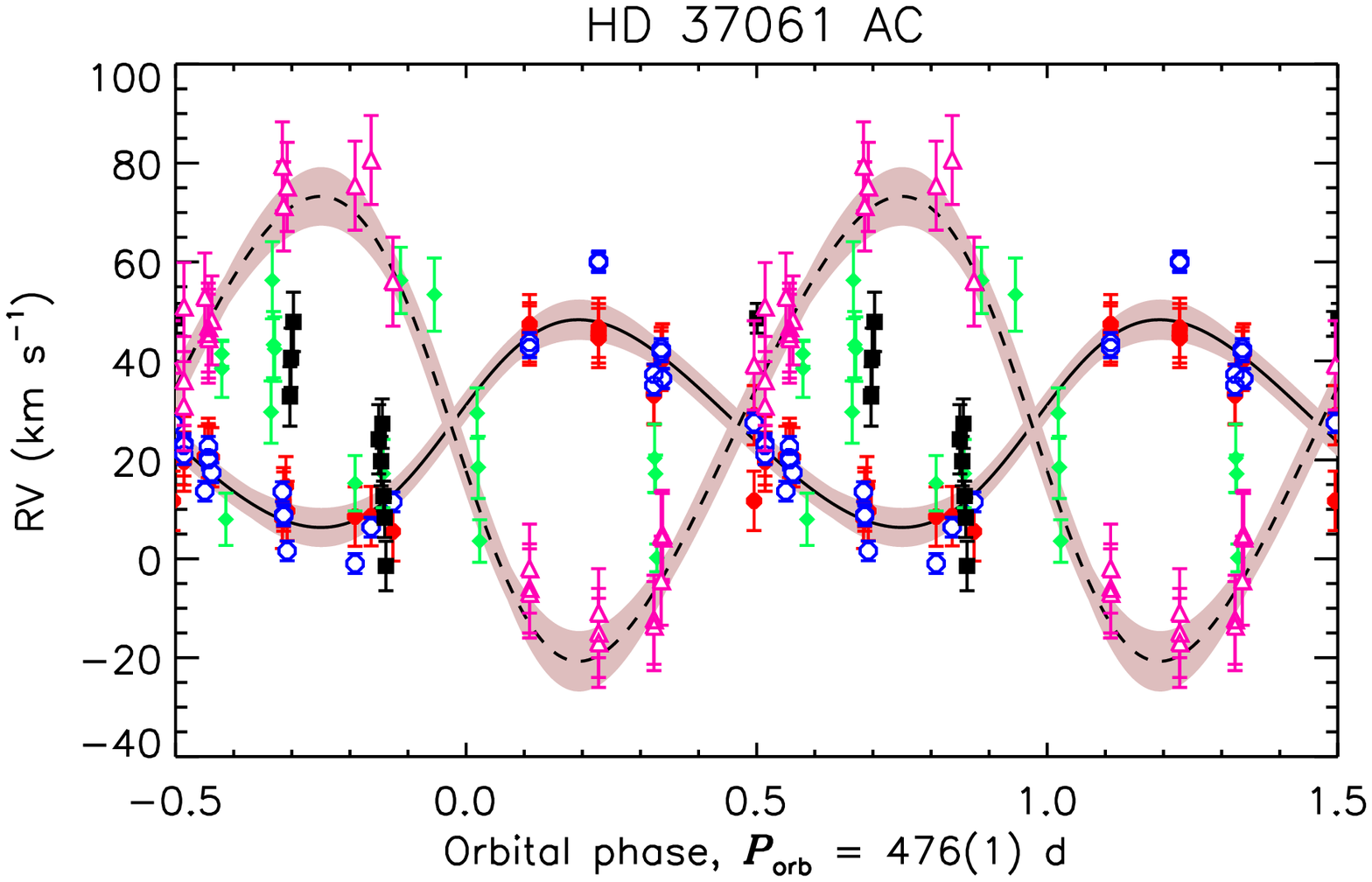}
   \includegraphics[width=\hsize]{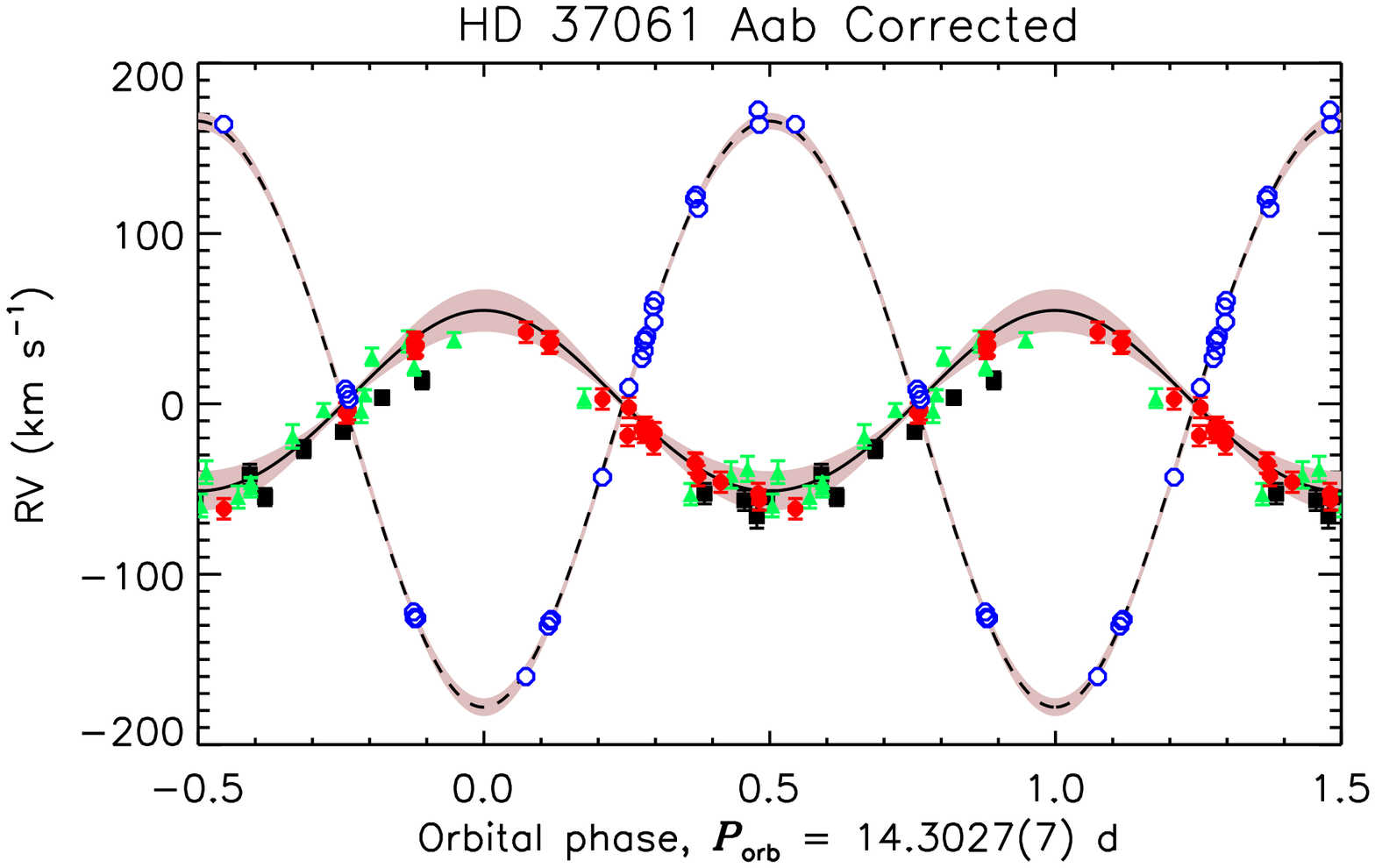}
      \caption[]{Radial Velocity curves. Open blue circles indicate Ab; filled red circles Aa; filled black squares the \protect\cite{1991ApJS...75..965M} measurements; filled green triangles the \protect\cite{1991ApJ...367..155A} measurements; and open purple triangles the C RVs. {\em Top}: Aab RVs, folded with the Aab orbital period, without correcting for the AC orbit. Solid (dashed) curves show the best-fit orbital models for the primary (secondary), shaded regions indicate 1$\sigma$ uncertainties from MCMC modeling. {\em Middle}: residual Aab RVs (after removal of orbital variation in top panel), and C RVs, folded with the AC orbital period. Solid and dashed curves are as in the top panel, but for A and C, respectively. {\em Bottom}: Aab RVs, corrected for the AC orbit, folded with Aab orbital period. Note that there is substantially less scatter as compared to the top panel.}
         \label{rv_curves}
   \end{figure}

Lomb-Scargle statistics were utilized in order to determine orbital periods from the RV measurements, using {\sc period04} \citep{2005CoAst.146...53L}. Periodograms are shown in Fig.\ \ref{Aab_periods}. Period analysis of the Ab component's RVs (bottom panel of Fig.\ \ref{Aab_periods}) yields maximum amplitude at 14.305(2)~d, where the number in brackets gives the uncertainty in the final digit. The same analysis of the Aa component's He~{\sc ii} 468.6 nm RVs yields 14.295(16)~d, which is compatible with the result for Ab and is most naturally interpreted to imply that these components are in orbit around one another. 

\cite{1991ApJS...75..965M} published 10 RV measurements of the Aa component, from which they determined a period of approximately 8 d. \cite{1991ApJ...367..155A} published a further 18 RVs, from which they determined a period of 19.139(3) d. Combining these measurements with our own yields 14.3027(7)~d, which is compatible with the value determined from the Ab component's RVs. There is no power at 19 d in the combined periodogram, thus the \citeauthor{1991ApJ...367..155A} period is not supported by our measurements. The Aa and Ab RVs are shown phased with the 14.3027 d period in the top panel of Fig.\ \ref{rv_curves}. The amplitude of the Aa RVs is consistent between our measurements and those published by \cite{1991ApJ...367..155A} and \cite{1991ApJS...75..965M}. As expected given the large difference in flux between Aa and Ab, the RV amplitude of Ab is much greater than that of Aa. There is also a large degree of scatter in the Aab RVs, outside of the formal uncertainties.

The bottom panel of Fig.\ \ref{AB_periods} shows the periodogram obtained for the C component's RVs, which yields a peak at 479(12)~d. This is consistent with the lack of RV variation over short timescales (see Table \ref{rvtab}), as well as the lower RV amplitude of the C component as compared to Aab. 

To see if the scatter in the Aab RVs may be related to the presence of the C component, residual RVs for Aa and Ab were obtained by fitting sinusoids using

\begin{equation}\label{sinfit}
{\rm RV} = {\rm RV}_0 + {\rm RV}_1 \sin{(\phi + \Phi_0)},
\end{equation}

\noindent where $\phi$ is the orbital phase of the Aab subsystem, and then subtracting the fit from the measurements (leaving out the ${\rm RV}_0$ component so as to leave the systemic velocity unchanged). Period analysis of the total residual Aa RVs (top panel of Fig.\ \ref{AB_periods}) yields a peak at 469(2)~d, consistent with the maximum-amplitude peak in the  C RVs; limiting the dataset to the ESPaDOnS data yields 484(18)~d. Residual Ab RVs yield maximum amplitude at 479(4)~d. The interferometric data analyzed below (\S~\ref{subsec:interf_orbit}) yields a period of 476.5(1.2)~d, with a 468~d period being firmly excluded. Since the shorter period only appears with inclusion of the literature data, it is likely an artifact caused by systematics in the older dataset arising from e.g.\ the contribution of the C component to the RV (which would have affected the gaussian fits used to measure the RVs). 

Residual RVs, and RV measurements of C, are shown phased with the 476~d period in the middle panel of Fig.\ \ref{rv_curves}. Residual Aa and Ab RVs vary in phase with one another, in antiphase with the C RVs, and the amplitude of the C RVs is additionally about twice that of the residual Aa and Ab RVs. This further indicates that the `scatter' in the Aab RVs is a consequence of the orbit of Aab and C about a common centre of mass. The residual literature RVs are not phased coherently with the 476~d period; the quality of the phasing is furthermore not greatly improved by adoption of the 468~d period. 

The bottom panel of Fig.\ \ref{rv_curves} shows the Aab RVs after correction for the variation in the Aab sub-system's centre of mass, via subtraction from the RVs of the best-fit sinusoid to the residual RVs. This correction reduces the scatter in the ESPaDOnS RVs to a magnitude similar to the formal uncertainties. 

\subsection{Orbital modeling}\label{subsec:orb}

\begin{table}
\centering
\caption{Orbital parameters for the Aab and AC sub-systems. For AC, $M_1$ refers to the mass of the combined Aab subsystem.}
\label{orbtab}
\begin{tabular}{l c c}
\hline
\hline
\\
Parameter & Aab & AC \\
\hline
$P_{\rm orb}$ (d) & 14.3027(7) & 476(1) \\
$T_0$ & 2440578.5(5) & 2453639(7) \\
$v_0$~(\kms) & $26 \pm 6$ & $28 \pm 3$ \\
$K_1$~(\kms) & $50 \pm 8$ & $21 \pm 4$ \\
$K_2$~(\kms) & $172 \pm 3$ & $47 \pm 6$ \\
$e$ & $<0.02$ & $0.09 \pm 0.06$ \\
$\omega~(^\circ)$ & -- & $100 \pm 5$ \\
$M_1 / M_2$ & $3.3 \pm 0.6$ & $2.2 \pm 0.5$ \\
$M_1\sin^3{i}$~(\msun) & $12.6 \pm 1.2$ & $9.4 \pm 3.2$ \\
$M_2\sin^3{i}$~(\msun) & $3.5 \pm 0.9$ & $4.5 \pm 1.4$ \\
$(M_1+M_2)\sin^3{i}$~(\mdot) & $16 \pm 2$ & $13 \pm 5$ \\
$a\sin{i}$~(AU) & $0.29 \pm 0.01$ & $2.8 \pm 0.3$ \\
\hline
\hline
\end{tabular}
\end{table}

For fitting the Aab RVs, the full dataset including literature measurements were used, with the RVs corrected for the orbital motion of the C component. For fitting the AC RVs, only the ESPaDOnS data were utilized, since the literature measurements are not phased coherently with the period. 

Orbital properties were determined using a Markov Chain Monte Carlo (MCMC) algorithm, with the radial velocity semi-amplitudes $K_1$ and $K_2$, the systemic velocity $v_0$, the eccentricity $e$, the argument of periapsis $\omega$, and the epoch $T_0$ as free parameters. The initial guess for $T_0$ is determined via sinusoidal fits to the RVs and residual RVs of the Aa component, with $T_0$ taken as the time of maximum RV in the cycle immediately preceding the first observation in the time series. The parameter space was explored by 20 independent Markov chains, starting from randomized initial conditions, using a synthetic annealing process that varies the parameter step size for test points by taking it to be the $\chi^2$-weighted standard deviation of the preceding (accepted) test points. The algorithm terminates when the Gelman-Rubin convergence condition is satisfied, i.e.\ the $\chi^2$ variance within chains is 10\% of the variance between chains \citep{1992StaSc...7..457G}. Once the algorithm has converged, fitting parameters and their uncertainties are derived from the peaks and standard deviations of their posterior probability density functions (PDFs). Derived parameters (mass ratios, projected masses, and semi-major axes) were obtained directly from the PDFs of the accepted test points. Mass ratios were obtained from RV semi-amplitudes, and projected masses and semi-major axes were obtained by applying Kepler's laws to the set of accepted test points.  

RV curves derived from orbital models are shown in Fig.\ \ref{rv_curves}, with the shaded regions indicating the 1$\sigma$ uncertainties. Orbital model parameters are provided in Table \ref{orbtab}. Modeling favours a circular orbit for Aab ($e < 0.02$), and a mildly eccentric orbit for AC ($e = 0.09 \pm 0.06$). 

\subsection{Results from interferometry}\label{subsec:interf_orbit}

\begin{table}
\centering
\caption{Gaia parallaxes and distances for sources within 30 arcseconds of NU Ori \protect\citep{2016A&A...595A...2G,2018A&A...616A...1G}.}
\label{disttab}
\begin{tabular}{l c c}
\hline
\hline
Source & $\pi$ & d \\
       & (mas) & pc \\
\hline
KPM2006 215 & $2.6 \pm 0.1$ & $385 \pm 15$ \\
KPM2006 216 & $2.7 \pm 0.3$ & $370 \pm 40$ \\
V* V2509 Ori  & $3.3 \pm 0.2$ & $300 \pm 20$ \\
KPM2006 219 & $3.0 \pm 0.3$ & $330 \pm 20$ \\
COUP 1500     & $2.44 \pm 0.06$ & $410 \pm 10$ \\
KPM2006 220 & $2.56 \pm 0.07$ & $390 \pm 10$ \\
KPM2006 217 & $2.6 \pm 0.1$ & $385 \pm 15$ \\
2MASS J05353229-0516269 & $1.7 \pm 0.4$ & $590^{+180}_{-110}$ \\
V* V1294 Ori & $2.66 \pm 0.04$ & $370 \pm 6$ \\   
\hline
\hline
\end{tabular}

\end{table}

\begin{figure*}  \centering 
  \includegraphics[width=18.cm, trim={0 8cm 0 0}]{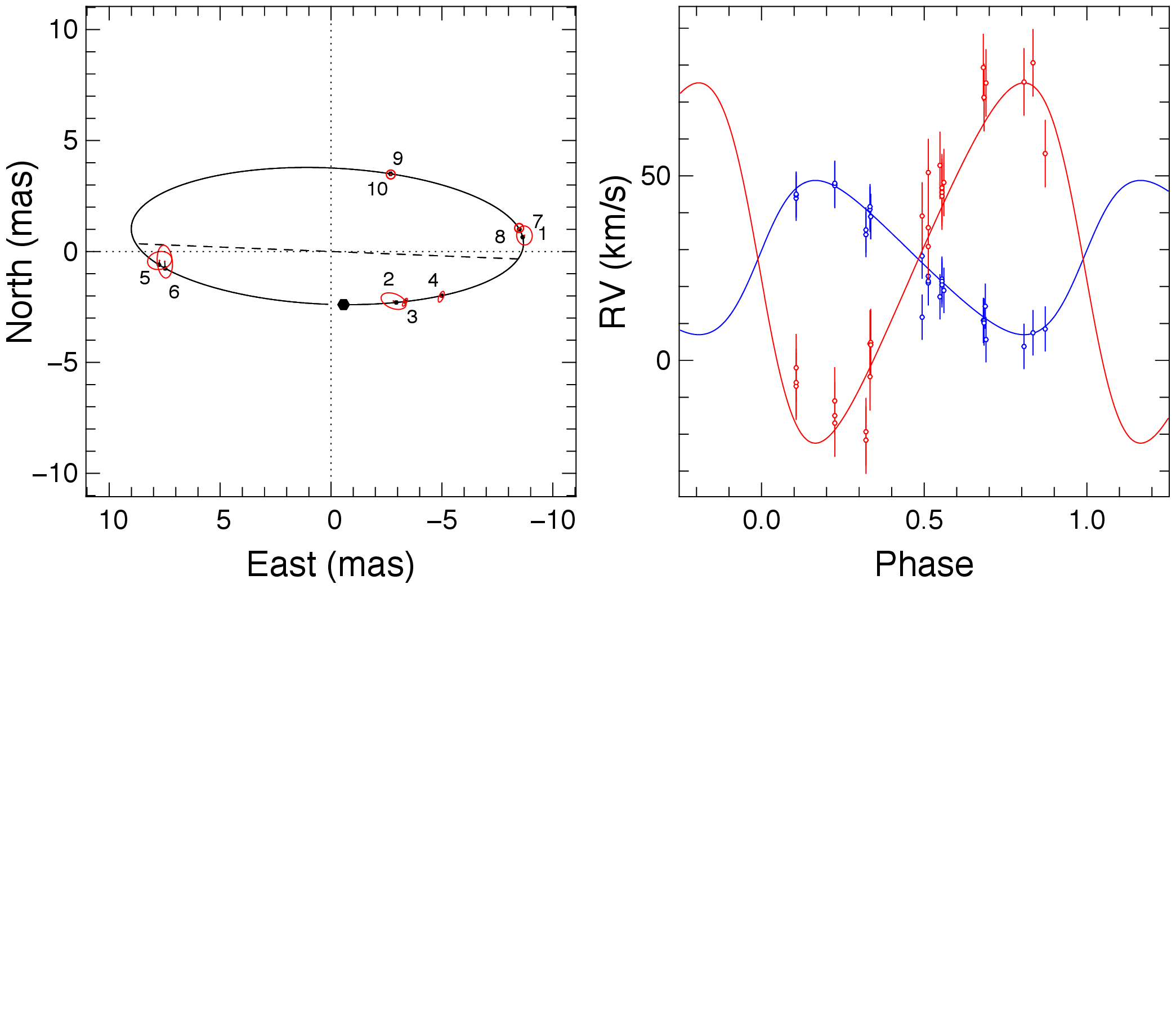}
  \caption{Best fit orbital solutions to the astrometric and velocimetric observations, assuming the distance of $370\pm30$\,pc. Left: motion of the secondary around the primary. Labels indicate the observation numbers corresponding to Table \ref{tab:interfero}. The periastron of the secondary is represented by a large filled hexagon and the line of nodes by a dashed line. Small black hexagons indicate the model positions of the C component, red dots indicate the measured positions, and red ellipses indicate the error ellipses. Right: radial velocities of the primary (blue) and the secondary (red).}
\label{fig:ABfit_GaiaAround}
\end{figure*}

\begin{table}\centering
\caption{Best-fit orbital parameters for the AC pair considering the interferometric observations and the SB2 radial velocities, with the distance fixed to the cluster distance inferred from Gaia parallaxes ($370\pm30$\,pc, first column), the distance as a free parameter (second column), and the distance inferred from the Gaia parallax of NU Ori (third column).} \label{tab:ABfit_GaiaAround}
 \begin{tabular}{lccc}
 \hline\hline\noalign{\smallskip}
 Parameter & Value & Value & Value \\
 \noalign{\smallskip}\hline\noalign{\smallskip}
 d (pc)    & $370 \pm 30$ & $348 \pm 17$ & $524 \pm 11$  \\
           & (fixed, cluster) & (free) & (fixed, Gaia) \\
 $T_0$ (MJD)  &  $53639(7)$  & 53640(8) & 53636(12) \\
 $P_{\rm orb}$ (d)  &  $476.5(1.2)$ & 476.5(1.3) & 476.8(2.1) \\
 $a$ (mas)  &  $9.06 \pm 0.17$  & $9.08 \pm 0.14$ & $8.89 \pm 0.15$ \\
 $e$   &  $0.226 \pm 0.025$  & $0.225 \pm 0.023$ & $0.227 \pm 0.022$ \\
 $\Omega$ ($^\circ$)  &  $87.7 \pm 1.5$ & $87.7 \pm 1.2$ & $87.7 \pm 1.2$  \\
 $\omega$ ($^\circ$)  &  $95.4 \pm 3.0$  & $95.5 \pm 3.0$ & $94.5 \pm 4.2$ \\
 $i_{\rm orb}$ ($^\circ$)  &  $70.1 \pm 0.9$ & $70.2 \pm 0.8$ & $69.5 \pm 0.8$ \\
 $K_{\rm A}$ (\kms)  &  $22.5 \pm 2.2$ & $20.8 \pm 1.6$ & $30.5 \pm 1.6$ \\
 $K_{\rm C}$ (\kms)  &  $52.1 \pm 4.1$ & $48.9 \pm 2.6$ & $71.6 \pm 2.4$ \\
 $v_0$ (\kms)  &  $27.5 \pm 1.1$ & $27.4 \pm 0.9$ & $27.6 \pm 0.9$ \\
 \noalign{\smallskip}\hline\noalign{\smallskip}
 $a_p$ (AU) & $ 3.38 \pm 0.26$ & $3.16 \pm 0.20$ & $4.66 \pm 0.18$ \\
 $M_{\rm A}$ (${M}_\odot$) & $16.2 \pm 3.8$ & $13.1 \pm 2.5$ & $41.6 \pm 3.2$ \\
 $M_{\rm C}$ (${M}_\odot$) & $7.0 \pm 1.7$ & $5.6 \pm 0.8$ & $17.7 \pm 1.5$ \\
 $M_{\rm A}/M_{\rm C}$  & $2.3 \pm 0.2$ & $2.3 \pm 0.2$ & $2.4 \pm 0.4$ \\
 \noalign{\smallskip}\hline

 \end{tabular}
\end{table}

We fit the interferometric observables with a simple binary model, representing the AC pair. The inner pair Aab is largely unresolved even by the longest baseline of VLTI. The remaining free parameters are the separation, the position angle, and the flux ratio between the secondary and the primary. The best fit is considered constant over one photometric band. The best-fit flux ratios are consistent among the different epochs ($0.190\pm0.015$ in H-band and $0.185\pm0.011$ in K-band). Inferred positions are summarized in Table~\ref{tab:interfero}, and shown in the left panel of Fig.\ \ref{fig:ABfit_GaiaAround}.

We simultaneously fit the resolved astrometric positions, the mean Aab residual RVs, and the C RVs. The residual RVs obtained from the literature were discarded as a solution could not be determined using these data. We followed the same convention for the orbital elements as detailed by \citet{2017A&A...601A..34L}. The uncertainties were estimated by fitting an ensemble of input datasets that followed the input mean and uncertainties, and computing the standard deviation of the best-fit parameters over this ensemble.

The Gaia parallax \citep{2016A&A...595A...2G,2018A&A...616A...1G}\footnote{Obtained from http://gea.esac.esa.int/archive/.} is $\pi = 1.91 \pm 0.04$~mas, implying a distance of $520 \pm 10$~pc, around 100 pc further than the usual distance from the ONC. Using the Gaia distance, the fit returns a mass for the Aab pair of 41 \msun, i.e.\ Aa should be an O-type star with \teff~$\sim 45$~kK. 

Given this implausible result, it seems possible that the Gaia astrometry for NU Ori is unreliable, possibly due to the influence of the C component being unaccounted for. Since NU Ori Aa is the central ionizing star of the M 43 H~{\sc ii} region, it must be physically associated with the ONC. Therefore, we examined the Gaia parallaxes of the 9 stars within 30'' of NU Ori, under the assumption that these are likely to also be members of the ONC. These are listed in Table \ref{disttab}. The mean and standard deviation for the dataset are $d = 390 \pm 80$~pc. One of the sources, a 2MASS object, is much further away (590 pc), and is likely a background source; excluding it, the mean and standard deviation are $370 \pm 30$ pc. This is close to the distance found by \cite{2008MNRAS.386..261M} using main sequence models ($390 \pm 10$~pc), as well as distances to non-thermal ONC sources determined using very long baseline interferometry ($390 \pm 20$~pc, \citealt{2007ApJ...667.1161S}; $388 \pm 5$~pc, \citealt{2017PhDT........55K}). We therefore imposed a distance of $370\pm30$\,pc. 

The best-fit orbit using the Gaia cluster distance is represented in Fig.~\ref{fig:ABfit_GaiaAround} and the corresponding orbital elements are summarised in Table~\ref{tab:ABfit_GaiaAround}. The quality of the fit is only marginally improved when removing the constraint on the distance. The $\chi^2_r$ decreases from 0.57 to 0.56. The best fit favours a slightly lower distance ($350\pm18$\,pc) and masses ($M_{\rm A} = 13.1\pm 2.5~{M}_\odot$ and $M_{\rm C} = 5.6\pm 0.8~{M}_\odot$). As can be seen from Table \ref{tab:ABfit_GaiaAround}, fits utilizing the cluster distance and with distance as a free parameter yield orbital and derived stellar parameters that overlap within uncertainty. The same is not true using the Gaia parallax for NU Ori itself. In this case, the derived masses are much higher, due to the much greater radial velocity semi-amplitudes necessary to reconcile the RV curves with the astrometry. The synthetic RV curves obtained using the Gaia distance are furthermore a noticeably poorer fit to the measured RVs, again suggesting that the parallax is likely in error. However, the Gaia distance solution still yields almost identical values for the eccentricity, the argument of periapsis, the systemic velocity, and the orbital inclination, indicating that these parameters are quite robust against any future revision in the distance. 

If we utilize the semi-major axis in AU obtained from fitting the RVs alone (Table \ref{orbtab}), and the orbital inclination and projected semi-major axis in mas from the interferometric data, we obtain a distance of $327\pm41$~pc, which is compatible with either the cluster distance or the best-fit distance where distance is left as a free parameter.

The mass obtained for the A component, $M_{\rm A} = 16 \pm 4$~\msun, is almost identical to the total projected mass obtained by modeling the Aab RVs, $M_{\rm A}\sin^3{i} = 16 \pm 2$~\msun. This suggests that the orbital axis of the Aab subsystem must have a similarly large inclination to that obtained for AC. 

\section{Magnetometry}\label{sec:mag}

   \begin{figure*}
   \centering
   \includegraphics[width=18.5cm]{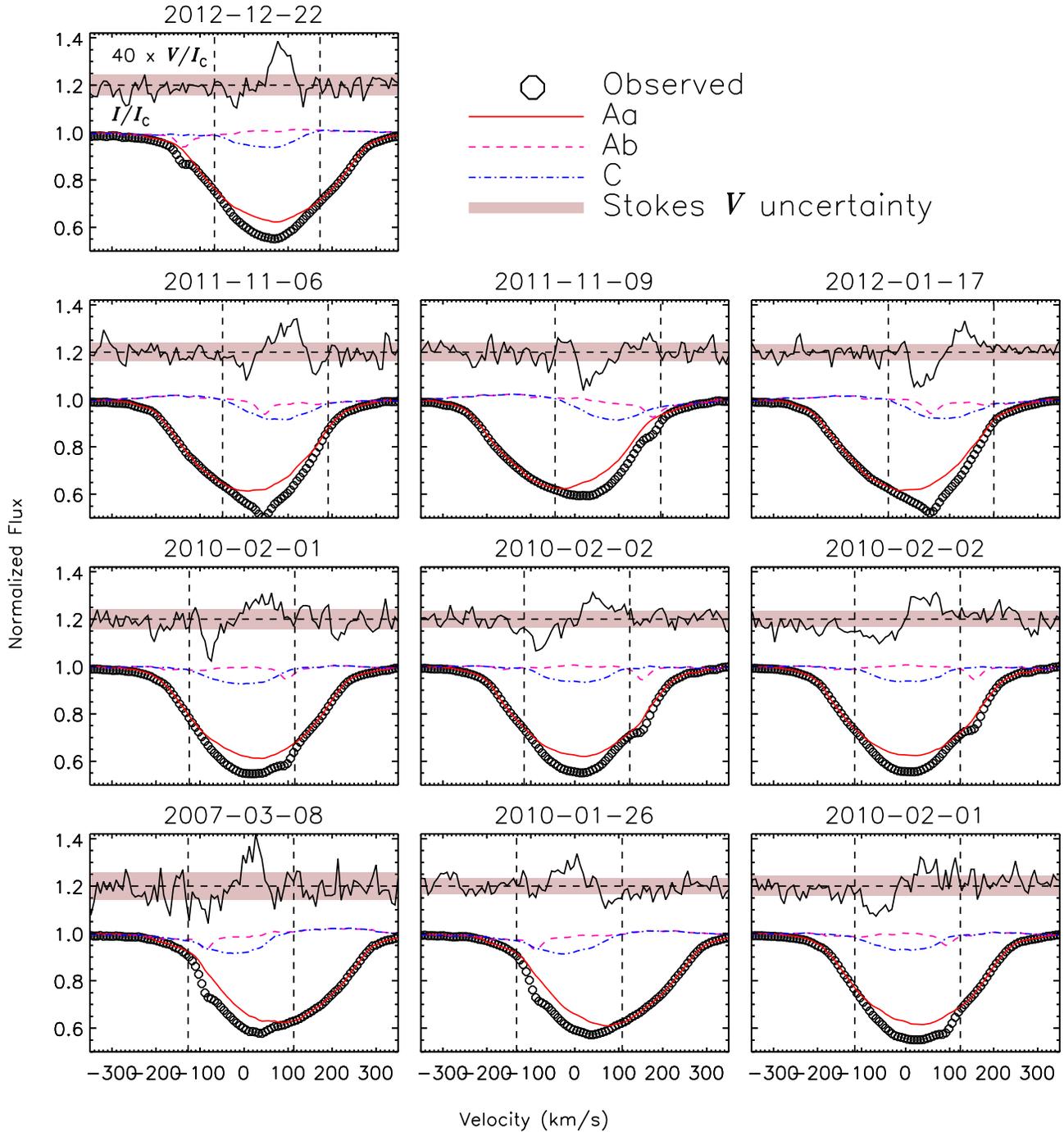}
      \caption[]{LSD Stokes $I$ (bottom) and $V$ (top) profiles yielding definite detections. The Stokes $V$ continuum is shown by a horizontal dashed line. Vertical dashed lines show the integration ranges for the C component. Note that the Stokes $V$ signature is confined within the line profile of the C component in all observations.}
         \label{NUOri_lsd}
   \end{figure*}

\begin{table*}
\centering
\caption{\bz~and \nz~measurements. Detection flags for Stokes $V$ (DF$_V$) or $N$ (DF$_N$) correspond to definite detections (DD), marginal detections (MD), or non-detections (ND).}
\resizebox{18.5 cm}{!}{
\label{bztab}
\begin{tabular}{l r r r r r r r r r r r r}
\hline
\hline
\\
           & \multicolumn{4}{c}{Aa} & \multicolumn{4}{c}{Ab} & \multicolumn{4}{c}{C} \\
           & \multicolumn{4}{c}{30 kK, met} & \multicolumn{4}{c}{17 kK, met} & \multicolumn{4}{c}{25 kK, met + He} \\
HJD        & \bz  & DF$_V$ & \nz  & DF$_N$ & \bz  & DF$_V$ & \nz    & DF$_N$ & \bz    & DF$_V$ & \nz    & DF$_N$ \\ 
-2450000   & (G) &        & (G) &        & (G) &        &   (G) &        &   (G) &        &   (G) &        \\
\hline
3747.87907 &  -221$\pm$  253 & ND &   305$\pm$  253 & ND &   -48$\pm$  203 & ND &   161$\pm$  204 & ND &  -122$\pm$  390 & ND &  -378$\pm$  390 & ND \\
3747.92079 &  -113$\pm$  250 & ND &   382$\pm$  250 & ND &    37$\pm$  264 & ND &  -256$\pm$  265 & ND &   316$\pm$  384 & ND &   185$\pm$  384 & ND \\
3747.96263 &  -375$\pm$  231 & ND &   563$\pm$  231 & ND &   -24$\pm$  170 & ND &    42$\pm$  170 & ND &  -200$\pm$  364 & ND &   548$\pm$  364 & ND \\
4167.72958 &  -164$\pm$  230 & ND &    69$\pm$  230 & ND &    69$\pm$  165 & ND &   -37$\pm$  165 & ND & -1222$\pm$  340 & {\bf DD} &   -11$\pm$  340 & ND \\
4167.76959 &  -250$\pm$  218 & ND &   298$\pm$  218 & ND &  -217$\pm$  160 & ND &     9$\pm$  160 & ND &  -806$\pm$  326 & ND &   830$\pm$  326 & ND \\
4167.81034 &  -252$\pm$  254 & ND &   -77$\pm$  254 & ND &    27$\pm$  137 & ND &   109$\pm$  137 & ND & -1641$\pm$  356 & MD &   -90$\pm$  356 & ND \\
5222.77964 &  -254$\pm$  183 & ND &    -3$\pm$  183 & ND &  -210$\pm$  146 & ND &    -1$\pm$  144 & ND &   126$\pm$  196 & ND &  -144$\pm$  196 & ND \\
5222.81932 &   -84$\pm$  182 & ND &    70$\pm$  182 & ND &   -61$\pm$  134 & ND &   -77$\pm$  135 & ND &   538$\pm$  207 & {\bf DD} &   298$\pm$  207 & ND \\
5228.75094 &  -250$\pm$  219 & ND &   -43$\pm$  219 & ND &   -70$\pm$  196 & ND &  -357$\pm$  198 & ND & -2342$\pm$  305 & {\bf DD} &  -377$\pm$  304 & ND \\
5228.79260 &  -193$\pm$  214 & ND &  -267$\pm$  214 & ND &  -159$\pm$  149 & ND &   167$\pm$  149 & ND & -1536$\pm$  272 & {\bf DD} &   242$\pm$  272 & ND \\
5229.79356 &  -247$\pm$  182 & ND &    -7$\pm$  182 & ND &    55$\pm$  114 & ND &   144$\pm$  114 & ND & -2504$\pm$  272 & {\bf DD} &   816$\pm$  271 & ND \\
5229.83319 &  -378$\pm$  176 & ND &  -153$\pm$  176 & ND &   206$\pm$  149 & ND &    -4$\pm$  148 & ND & -2369$\pm$  268 & {\bf DD} &    -2$\pm$  267 & ND \\
5871.10842 &   814$\pm$ 1205 & ND &  -520$\pm$ 1205 & ND &   371$\pm$ 1198 & ND &  -647$\pm$ 1201 & ND &   552$\pm$ 1254 & ND &  1297$\pm$ 1255 & ND \\
5872.08868 &  -151$\pm$  229 & ND &   596$\pm$  229 & ND &    -7$\pm$  156 & ND &  -204$\pm$  156 & ND &  -135$\pm$  254 & ND &   572$\pm$  254 & ND \\
5872.15256 &   178$\pm$  200 & ND &  -352$\pm$  200 & ND &   -42$\pm$  166 & ND &   -15$\pm$  166 & ND &  -378$\pm$  211 & {\bf DD} &   226$\pm$  211 & ND \\
5875.03292 &  -296$\pm$  203 & ND &   164$\pm$  203 & ND &    11$\pm$  137 & ND &    34$\pm$  137 & ND &  -555$\pm$  211 & {\bf DD} &   -94$\pm$  211 & ND \\
5930.71891 &  -282$\pm$  263 & ND &  -159$\pm$  263 & ND &  -102$\pm$  200 & ND &   174$\pm$  201 & ND &  -881$\pm$  257 & ND &    90$\pm$  256 & ND \\
5943.90765 &   -34$\pm$  177 & ND &   -41$\pm$  177 & ND &   -22$\pm$  148 & ND &    19$\pm$  148 & ND & -1274$\pm$  198 & {\bf DD} &  -339$\pm$  198 & ND \\
5961.75237 &  -201$\pm$  169 & ND &  -283$\pm$  169 & ND &   -40$\pm$  153 & ND &    15$\pm$  153 & ND &   324$\pm$  186 & ND &   -85$\pm$  186 & ND \\
6257.94322 &  -234$\pm$  909 & ND & -1411$\pm$  910 & ND &  -510$\pm$  617 & ND &   166$\pm$  614 & ND &  -280$\pm$  834 & ND &  -987$\pm$  834 & ND \\
6266.86057 &  2707$\pm$ 6075 & ND &   452$\pm$ 6075 & ND &   648$\pm$ 5792 & ND & -1940$\pm$ 5805 & ND & 11433$\pm$ 5901 & ND &  3395$\pm$ 5897 & ND \\
6266.90271 &   121$\pm$  768 & ND &  -546$\pm$  768 & ND &   387$\pm$  455 & ND &  -167$\pm$  454 & ND &  -401$\pm$ 1117 & ND &   361$\pm$ 1117 & ND \\
6266.93881 &  1093$\pm$  927 & ND &   426$\pm$  927 & ND &  -778$\pm$  688 & ND &   617$\pm$  685 & ND &   353$\pm$ 1300 & ND &   344$\pm$ 1300 & ND \\
6283.97259 &  -119$\pm$  227 & ND &   123$\pm$  227 & ND &    27$\pm$  130 & ND &   -16$\pm$  130 & ND & -1245$\pm$  403 & {\bf DD} &  -201$\pm$  402 & ND \\
6286.92096 &    83$\pm$  329 & ND &   546$\pm$  329 & ND &   288$\pm$  308 & ND &   -66$\pm$  307 & ND &   -92$\pm$  372 & ND &   370$\pm$  372 & ND \\
6286.95840 &  -214$\pm$  968 & ND &  -633$\pm$  968 & ND &   108$\pm$  903 & ND &   -38$\pm$  903 & ND &   302$\pm$ 1146 & ND &   974$\pm$ 1146 & ND \\
6286.99299 &    18$\pm$  262 & ND &    70$\pm$  262 & ND &   127$\pm$  227 & ND &  -232$\pm$  227 & ND &   197$\pm$  307 & ND &   176$\pm$  307 & ND \\
6289.78387 &   -96$\pm$  263 & ND &   113$\pm$  263 & ND &   -28$\pm$  196 & ND &   299$\pm$  197 & ND & -1580$\pm$  319 & MD &    33$\pm$  319 & ND \\
\hline
\hline
\end{tabular}
}
\end{table*}

   \begin{figure}
   \centering
   \includegraphics[width=8.5cm]{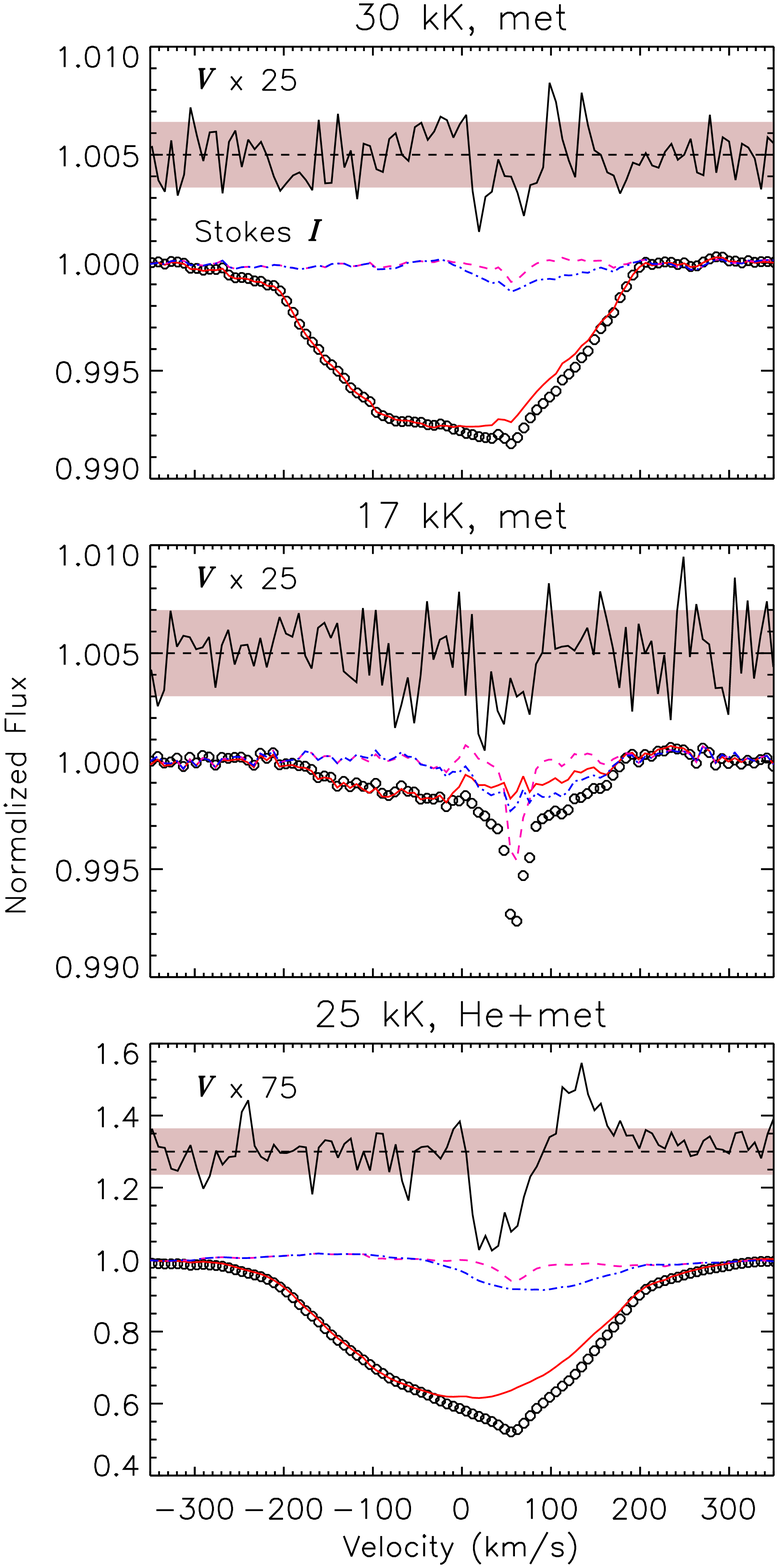}
      \caption[]{Comparison of LSD profiles extracted using line masks with different effective temperatures customized to emphasize the contribution from one of the 3 stars: 30 kK, Aa; 17 kK, Ab; 25 kK, C. Extracted and disentangled Stokes $I$ profiles are as in Fig.\ \ref{NUOri_lsd}. Note that the exclusion of He lines in the 30 kK and 17 kK line masks leads to much weaker Stokes $I$ profiles as compared to those obtained from the 25 kK mask, which includes He lines.}
         \label{nuori_mask_compare}
   \end{figure}

As a first step to analysis of NU Ori's magnetic field, least-squares deconvolution (LSD) profiles were extracted using a line mask developed from an extract stellar request from the Vienna Atomic Line Databsase \citep[VALD3;][]{piskunov1995,ryabchikova1997,kupka1999,kupka2000,2015PhyS...90e4005R} using the stellar parameters determined for NU Ori Aa (\teff~$=30.5 \pm 0.5$~kK, $\log{g} = 4.2 \pm 0.1$) by \cite{2011AA...530A..57S}. These parameters were selected since \cite{2008MNRAS.387L..23P} identified the Aa component as the magnetic star, given that the Stokes $V$ signature is much wider than the \vsini~of the secondary component. The line mask was cleaned of all H lines, as well as lines strongly blended with H line wings, lines in spectral regions strongly affected by telluric contamination, lines blended with nebular or interstellar features, and lines in spectral regions affected by ripples. While He lines are often removed due to substantial differences between magnetometry results obtained from He vs.\ metallic lines \citep[e.g.][]{2015MNRAS.447.1418Y,2015MNRAS.449.3945S,2018MNRAS.475.5144S}, in this case He lines were left in the mask since the majority of the Stokes $V$ line flux comes from these lines, and the Stokes $V$ profiles extracted using a line mask with He lines excluded did not result in detectable Zeeman signatures. Due to the very high \vsini~of the Aa component, LSD profiles were extracted using a velocity range of $\pm 600$~\kms~(in order to include enough continuum for normalization) and a velocity pixel size of 7.2~\kms, or 4 times the average ESPaDOnS velocity pixel (thus raising the per-pixel S/N by about a factor of 2). The significance of the signal in Stokes $V$ was evaluated using False Alarm Probabilities (FAPs), with observations classified as definite detections (DDs), marginal detections (MDs), or non-detections (NDs) according to the criteria described by \cite{1992AA...265..669D,d1997}.  Since FAPs essentially evaluate the statistical significance of the Stokes $V$ signal inside the stellar line by comparing it to the noise level, they are primarily senstive to the amplitude of Stokes $V$, which unlike \bz~is not strongly dependent on rotational phase. FAPs are thus a complementary means of checking for the presence of a polarization signal, the principal advantage being that they can detect a magnetic field even at magnetic nulls, i.e.\ \bz~$=0$.

Due to the presence of the two companion stars, the LSD profiles were disentangled using an iterative algorithm similar to the one employed by \cite{2006AA...448..283G}. This resulted in the unexpected discovery that it is most likely the C component, rather than the Aa component, that hosts the magnetic field. Fig.\ \ref{NUOri_lsd} shows the 10 LSD profiles yielding DDs in Stokes $V$, with the disentangled Aa, Ab, and C profiles. In all cases the Stokes $V$ signature is located entirely inside the line profile of the C component. In some observations, in which the Aa and C components have very different RVs (e.g. 2010-01-26, 2011-11-06, 2012-01-17), Stokes $V$ is clearly offset from the central velocity of Aa. This indicates that the identification of the primary as the magnetic star by \cite{2008MNRAS.387L..23P} was mistaken. 

In an effort to improve the quality of the LSD profiles, a second VALD3 line mask was obtained, this time with stellar parameters closer to those inferred for NU Ori C (\teff~$=24$~kK, $\log{g} = 4.25$; see \S~\ref{sec:stellar_parameters}). This line mask was cleaned as before, this time with the addition that lines obviously dominated by NU Ori Aa were excluded. Of the 923 lines in the original mask, 483 remained after cleaning. It is these LSD profiles that are shown in Fig.\ \ref{NUOri_lsd}. The detection flags obtained for Stokes $V$ and $N$ are summarized in Table \ref{bztab}. Only 10/28 observations yield DDs in Stokes $V$ for NU Ori C; 2 observations yield MDs; and the remainder are NDs. All $N$ profiles yield NDs, as expected for normal instrument operation. 

The longitudinal magnetic field \bz~\citep{mat1989} was evaluated by shifting the disentangled C profiles to their rest velocities, normalizing to the continuum in order for the Stokes $I$ equivalent width to be as accurate as possible, and using an integration range of $\pm 120$~\kms. \bz~measurements, and the analogous \nz~measurements obtained from the $N$ profiles, are reported in Table \ref{bztab}. 

Since the line profiles of the three components are strongly blended in all observations, clean measurements of Aa and Ab are difficult to obtain. In an effort to isolate the contributions of the different stellar components, line masks were prepared using the original 30 kK line mask (for Aa), and a 17 kK line mask (for Ab). In both cases, all He~{\sc i} lines were removed, since C contributes a significant amount of flux to these lines. The line masks were then cleaned to remove any lines with obviously dominant contributions from the other stars. The final line masks contained 111 lines (for the 30 kK mask), and 112 lines (for the 17 kK mask). Examples of the resulting LSD profiles are shown together with the disentangled Stokes $I$ profiles in Fig.\ \ref{nuori_mask_compare}. Since the metallic lines are typically much weaker than the He lines, the LSD profiles extracted with these masks have much shallower Stokes $I$ profiles than those obtained using the 25 kK metallic + He mask. 

The tailored line masks were somewhat successful in reducing the contributions to Stokes $I$ of the other stars, however in both cases there is still some residual influence. FAPs and \bz~were evaluated from the disentangled profiles using the same method as for those from the 25 kK line mask, with integration ranges appropriate to the star in question ($\pm 220$~\kms~for Aa, $\pm 20$~\kms~for Ab). All observations yield NDs in both Stokes $V$ and $N$. The \bz~measurements obtained from the 30 kK LSD profiles show a systematic bias towards negative values, indicating that the contribution of C to Stokes $V$ was not fully removed. No such bias is apparent in the \bz~measurements from the disentangled LSD profiles of Ab from the the 17 kK line mask. This is likely because the much narrower spectral lines of Ab are only blended with those of C at certain observations, unlike those of Aa, which are blended at all times.

\subsection{Rotational period}

   \begin{figure}
   \centering
   \includegraphics[width=8.5cm,trim=25 25 0 0]{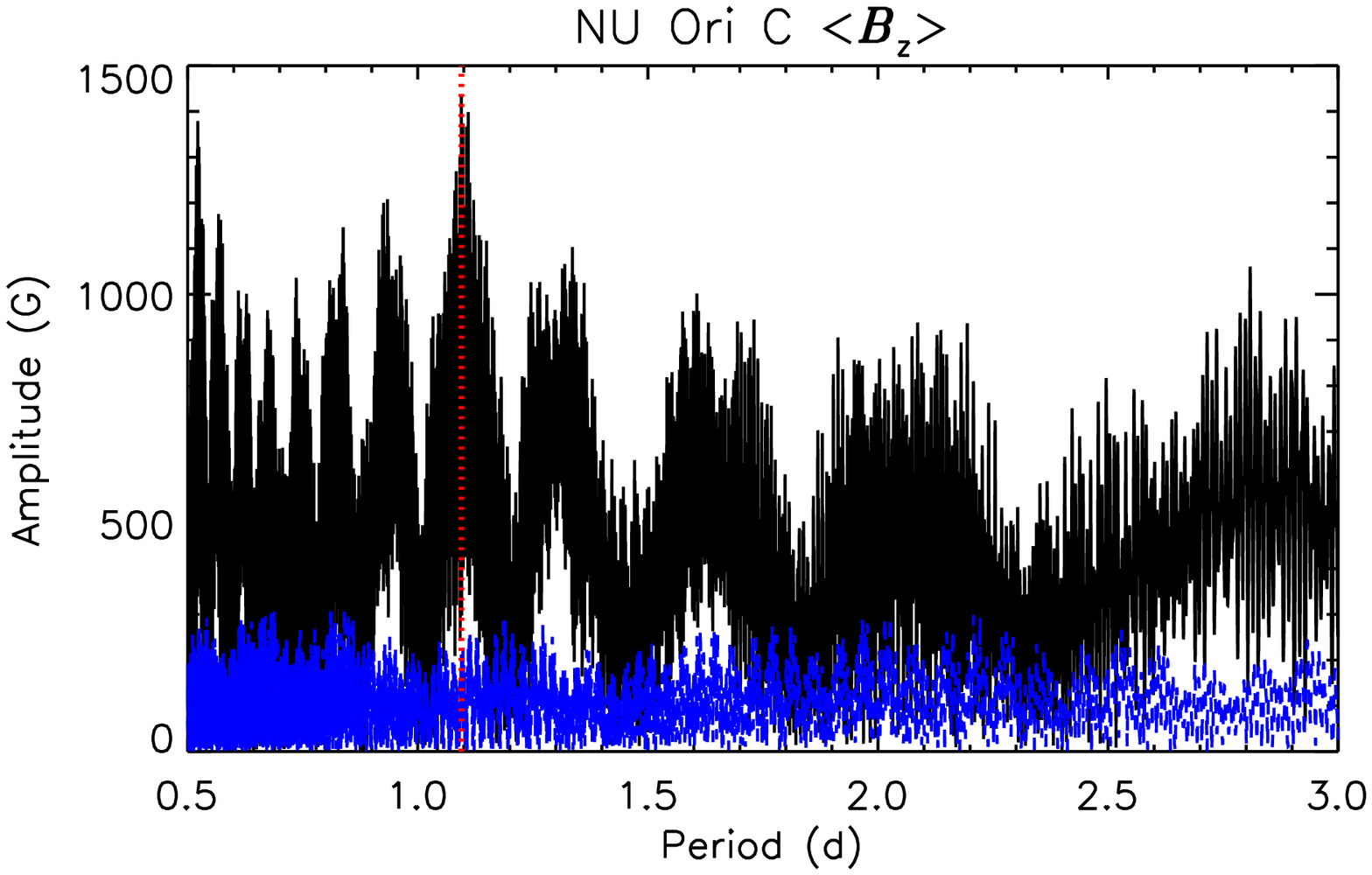}
      \caption[]{Periodogram for NU Ori C \bz~measurements (solid black) and \nz~measurements (dashed blue). The dotted red line indicates the maximum amplitude period.}
         \label{NUOri_B_bz_periods}
   \end{figure}

   \begin{figure}
   \centering
   \includegraphics[width=8.5cm]{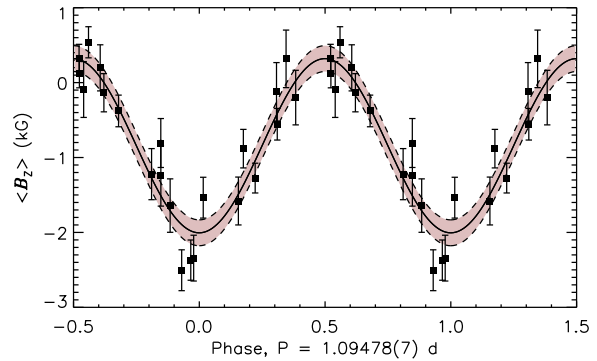}
      \caption[]{\bz~measurements of the NU Ori C from disentangled LSD profiles, folded with the rotational period. The solid line shows the best-fit first-order sinusoid, and the grey shaded regions indicate the 1$\sigma$ uncertainty in the fit.}
         \label{HD37061_B_bz}
   \end{figure}

NU Ori's rotational period was previously reported as 1.0950(4)~d by \cite{2018MNRAS.475.5144S}, however given the identification of NU Ori C rather than NU Ori Aa as the magnetic star, this should be revisited. Fig.\ \ref{NUOri_B_bz_periods} shows the periodogram for NU Ori C's \bz~and \nz~measurements. As expected, there is low amplitude in \nz~at all periods. The \bz~periodogram shows maximum amplitude at 1.09478(7) d, compatible with the period found by \cite{2018MNRAS.475.5144S} albeit at a higher precision. The greater precision is due first to the increased amplitude of the \bz~variation relative to the \bz~uncertainty as compared to the results obtained when Aa was assumed to be the magnetic star, and second to inclusion of the observations reported by \cite{2008MNRAS.387L..23P} (which were neglected by \citeauthor{2018MNRAS.475.5144S}). The FAP of this peak is 0.0005, which is much lower than the \nz~peak (0.21), and lower than the peak of 0.08 in the periodogram reported by \citeauthor{2018MNRAS.475.5144S}. This confirms the period found by \citeauthor{2018MNRAS.475.5144S}, but at a much higher degree of certainty.

The epoch $T_0 = 2453747.5(1)$ was determined by fitting a sinusoid to the data and determining the time of $|\langle B_z \rangle|_{\rm max}$ in the cycle immediately preceding the first observation. \bz~is shown phased with this ephemeris in Fig.\ \ref{HD37061_B_bz}. For the purposes of modeling the star's magnetic dipole, a sinusoidal fit was performed using the relation

\begin{equation}\label{bz_fit}
\langle B_z \rangle = B_0 + B_1 \sin{(\phi + B_2)},
\end{equation}

\noindent where $\phi$ is the rotational phase and $B_2$ is a phase offset. The fit and its uncertainties are shown in Fig.\ \ref{HD37061_B_bz}. The resulting coefficients are $B_0 = -0.85 \pm 0.07$~kG and $B_1 = 1.18 \pm 0.09$~kG. The reduced $\chi^2$ of the fit is 1.6. Fitting \bz~with the second harmonic yielded a reduced $\chi^2$ of 1.7, i.e.\ the fit is not improved, indicating that the star's \bz~variation is adequately described by a dipole. If the fit to \bz~{\em had} been improved by addition of a second harmonic, a more likely explanation than a multipolar field would be that the contributions of Aa and Ab had been inadequetly removed; since a second harmonic is unnecessary, this also suggests that the disentangling procedure was successful in isolating the components. 

\section{Discussion}\label{sec:discussion}

\begin{table}
\centering
\caption{Stellar, rotational, magnetic, and magnetospheric parameters of the  NU Ori components. Parameters obtained from the literature are indicated with superscripts corresponding to the following reference key: a) \protect\cite{petit2013}; b) \protect\cite{2011AA...530A..57S}; c) \protect\cite{2018ApJ...859..166F}. Surface magnetic dipole strengths for Aa and Ab correspond to 1/2/3$\sigma$ upper limits (see text).}
\resizebox{8.5 cm}{!}{
\label{phystab}
\begin{tabular}{l c c c}
\hline
\hline
\\
Parameter & Aa & Ab & C \\
\hline
$V$~(mag) & \multicolumn{3}{c}{6.83} \\
$d$~(pc) & \multicolumn{3}{c}{$370 \pm 30$} \\
$A_V$~(mag) & \multicolumn{3}{c}{$2.08 \pm 0.25$$^{a}$} \\
$M_{V}$~(mag) & $-3.1 \pm 0.2$ & $0.1 \pm 0.4$ & $-1.5 \pm 0.2$ \\
$BC$~(mag) & $-2.92 \pm 0.03$ & $-1.3 \pm 0.2$ & $-2.2 \pm 0.1$ \\
$M_{\rm bol}$~(mag) & $-6.2 \pm 0.2$ & $-2.0 \pm 0.7$ & $-4.3 \pm 0.7$ \\
$\log{(L/L_\odot)}$ & $4.29 \pm 0.09$ & $2.4 \pm 0.3$ & $3.3 \pm 0.2$ \\
$T_{\rm eff}$~(kK) & $30.5 \pm 0.5$$^{b}$ & $15.2 \pm 1.4$ & $22.2 \pm 1.0$ \\
$\log{g}$ & $4.2 \pm 0.1$$^{b}$ & $4.33 \pm 0.01$ & $4.28 \pm 0.02$ \\
$\log{(t_{\rm cl} / {\rm Myr})}$ & \multicolumn{3}{c}{$5.75 \pm 0.25$$^{c}$} \\
$\log{(t_{\rm HRD} / {\rm Myr})}$ & \multicolumn{3}{c}{$6.5 \pm 0.2$} \\
$M_*~(M_\odot)$ & $14.9 \pm 0.5$ & $3.9 \pm 0.7$ & $7.8 \pm 0.7$ \\
$R_*~(R_\odot)$ & $5.1 \pm 0.3$ & $2.2 \pm 0.2$ & $3.3 \pm 0.2$ \\
\vsini~(\kms) & $190 \pm 10$ & $10 \pm 5$ & $100 \pm 10$ \\
$v_{\rm mac}$~(\kms) & $20 \pm 10$ & $5\pm5$ & $5\pm5$ \\
$P_{\rm rot}$~(d) & 0.5--1.5 & 0.3--12.6 & 1.09478(7) \\
$T_0$ & -- & -- & 2455222.2(1) \\
$v_{\rm eq}$~(\kms) & -- & -- & $175 \pm 25$ \\
$R_{\rm K}~(R_*)$ & -- & -- & $2.6 \pm 0.1$ \\
$i_{\rm rot}~(^\circ)$ & -- & -- & $38 \pm 5$ \\
$\beta~(^\circ)$ & -- & -- & $62 \pm 6$  \\
$B_{\rm d}$~(kG) & $<0.23/0.49/0.95$ & $<0.1/0.6/6.0$ & $7.9 \pm 1.5$ \\
$R_{\rm A}~(R_*)$ & -- & -- & $18^{+5}_{-1}$ \\
$\log{R_{\rm A}/R_{\rm K}}$ & -- & -- & $0.84^{+0.09}_{-0.03}$ \\
\hline
\hline
\end{tabular}
}
\end{table}

\subsection{Stellar Parameters}\label{sec:stellar_parameters}

   \begin{figure}
   \centering
   \includegraphics[width=8.5cm]{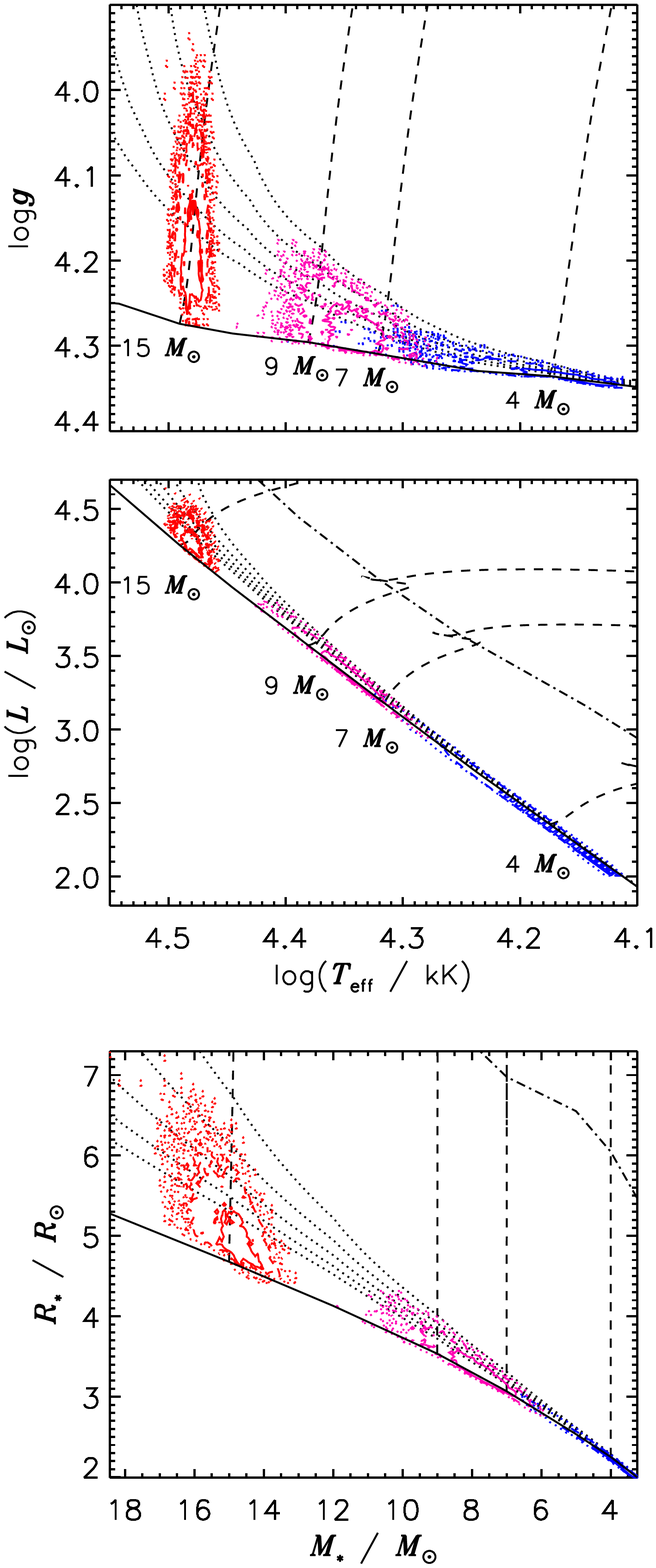}
      \caption[]{Stellar parameters of the components from Monte Carlo modelling. Contours indicate test point density (solid/dashed/dotted: 1/2/3$\sigma$), with red/purple/blue indicating Aa, C, and Ab. Dashed lines indicate evolutionary tracks from the rotating \protect\cite{ekstrom2012} models. Solid and dot-dashed lines show the zero-age and terminal-age main sequences. Dotted lines indicate isochrones from $\log{(t/{\rm yr})} = 6.5$ to 6.9 in increments of 0.1.} 
         \label{nuori_physpar_mc}
   \end{figure}

To obtain stellar parameters for the system components, a Monte Carlo algorithm was used similar to that described by \cite{pablo_epslup}. The algorithm works by populating the \teff-$\log{g}$ diagram with test points drawn from Gaussian distributions in \teff~and $\log{g}$, interpolating stellar parameters via evolutionary models and the orbital relationships of the components, and rejecting points that are inconsistent with known observables. Probability density maps for the three components are shown in Fig.\ \ref{nuori_physpar_mc} on the \teff$-\log{g}$ plane, the HRD, and the $R_*-M_*$ plane, and the derived physical parameters are listed in Table \ref{phystab}. 

Since the physical parameters of the Ab and C components are difficult to constrain directly from the spectrum, which is dominated by the Aa component, these were fixed using the mass ratios from Table \ref{orbtab}, following the assumption that close binaries are primordial and, hence, the stars must be coeval \citep{1994MNRAS.271..999B}. The evolutionary models of \cite{ekstrom2012}, including the effects of rotation, were utilized to determine ages and masses. Test points were accepted if they resulted in 1) projected masses $M_{\rm A}\sin^3{i}$ and $M_{\rm C}\sin^3{i}$ consistent with the inclination $i_{\rm orb} = 70.1 \pm 0.9^\circ$ determined from interferometric modeling (\S \ref{subsec:interf_orbit}), and 2) a combined absolute magnitude $M_V$ consistent with the system's observed $V$ magnitude, distance, and extinction (see Table \ref{phystab}). The Aa component's physical parameters, which were used to generate test points, were obtained from \cite{2011AA...530A..57S}, who used the NLTE {\sc fastwind} code to model the star's spectrum. While they did not account for contributions from the other two stars, their results were primarily sensitive to the strength of He~{\sc ii} lines, which should not be affected by Ab and C. 

The bolometric luminosity contributions of the three components derived from their mass ratios are 90\% for Aa, 1\% for Ab, and 9\% for C. As a sanity check, evaluation of the Planck function for the inferred effective temperatures of the 3 components indicates that the fractional flux contribution of the C component in the H and K bands should be around 20\%, agreeing well with the interferometric results (\S~\ref{subsec:interf_orbit}).

Solving the projected masses for Aab (Table \ref{orbtab}) using the masses inferred from the HRD yields $i_{\rm orb,Aab} = 72 \pm 9^\circ$ using $(M_{\rm Aa} + M_{\rm Ab}) \sin^3{i}$, $i_{\rm orb,Aa} = 72 \pm 6^\circ$ using $M_{\rm Aa}\sin^3{i}$, and $i_{\rm orb,Ab} = 74 \pm 24^\circ$ using $M_{\rm Ab}\sin^3{i}$, i.e.\ consistent with the orbital inclination of the AC sub-system determined by interferometry. This suggests that the orbital axes of the Aab and AC sub-systems are aligned. The masses of C and Aab, respectively $7.8 \pm 0.7$~\msun~and $18.8 \pm 1.2$~\msun, are consistent with the interferometric masses, albeit more precise by a factor of about 3 (although these results are more strongly model-dependent). 

\cite{1981AA....99..126H} showed that the timescales for spin-orbit alignment and pseudo-synchronization should be comparable, and much less than the timescale for circularization. There is no evidence for synchronization of the orbital and rotational periods of NU Ori C, which is consistent with its wide and eccentric orbit. As shown below in \S~\ref{sec:bfield}, the orbital and rotational axes of NU Ori C are most likely misaligned. 

The Aab orbit is nearly circular, and therefore may be expected to exhibit synchronized orbital and rotational motion as well as aligned orbital and rotational axes. Given the high \vsini~of Aa, it is impossible for the rotational and orbital periods to be perfectly synchronized, since the maximum rotation period (if the rotational axis is exactly perpendicular to the line of sight) is $1.36 \pm 0.15$~d, much less than the 14.3~d orbital period. Assuming spin-orbit alignment and adopting $i_{\rm rot}=70^\circ$ yields $P_{\rm rot}=1.27 \pm 0.14$~d. This is very close to $1/11^{th}$ of the orbital period, and could indicate a spin-orbit resonance. Assuming spin-orbit alignment for Ab yields $P_{\rm rot} = 10 \pm 6$~d, which is compatible (within the large uncertainty) with perfectly synchronized rotation, or with a rotation period of $1/2$ or $1/3^{rd}$ of the orbital period. While the the rotational properties of the Aab components are compatible with both spin-orbit alignment and pseudo-synchronization, this obviously cannot be confirmed. Given the system's youth \citep[about 500 kyr;][]{2018ApJ...859..166F}, there has not been much time for orbital evolution, and it may also be that it was born close to its current circular configuration. Alternatively, dynamical interactions with the C component may have circularized and hardened the Aab sub-system's orbit \citep{1979A&A....77..145M}.

\subsection{Magnetic Field \& Magnetosphere of NU Ori C}\label{sec:bfield}

   \begin{figure*}
   \centering
\begin{tabular}{cc}
   \includegraphics[width=8cm]{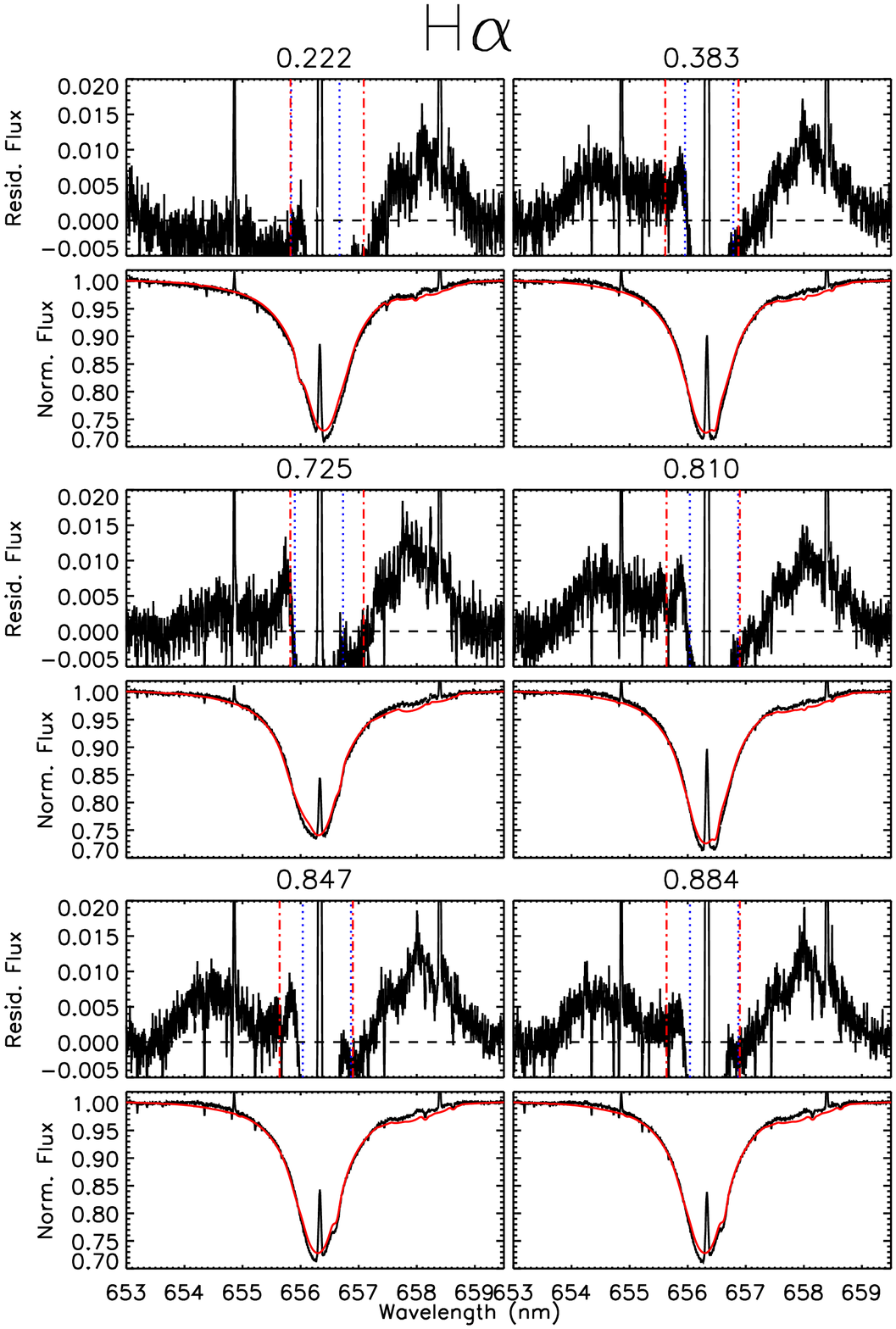} & 
   \includegraphics[width=8cm]{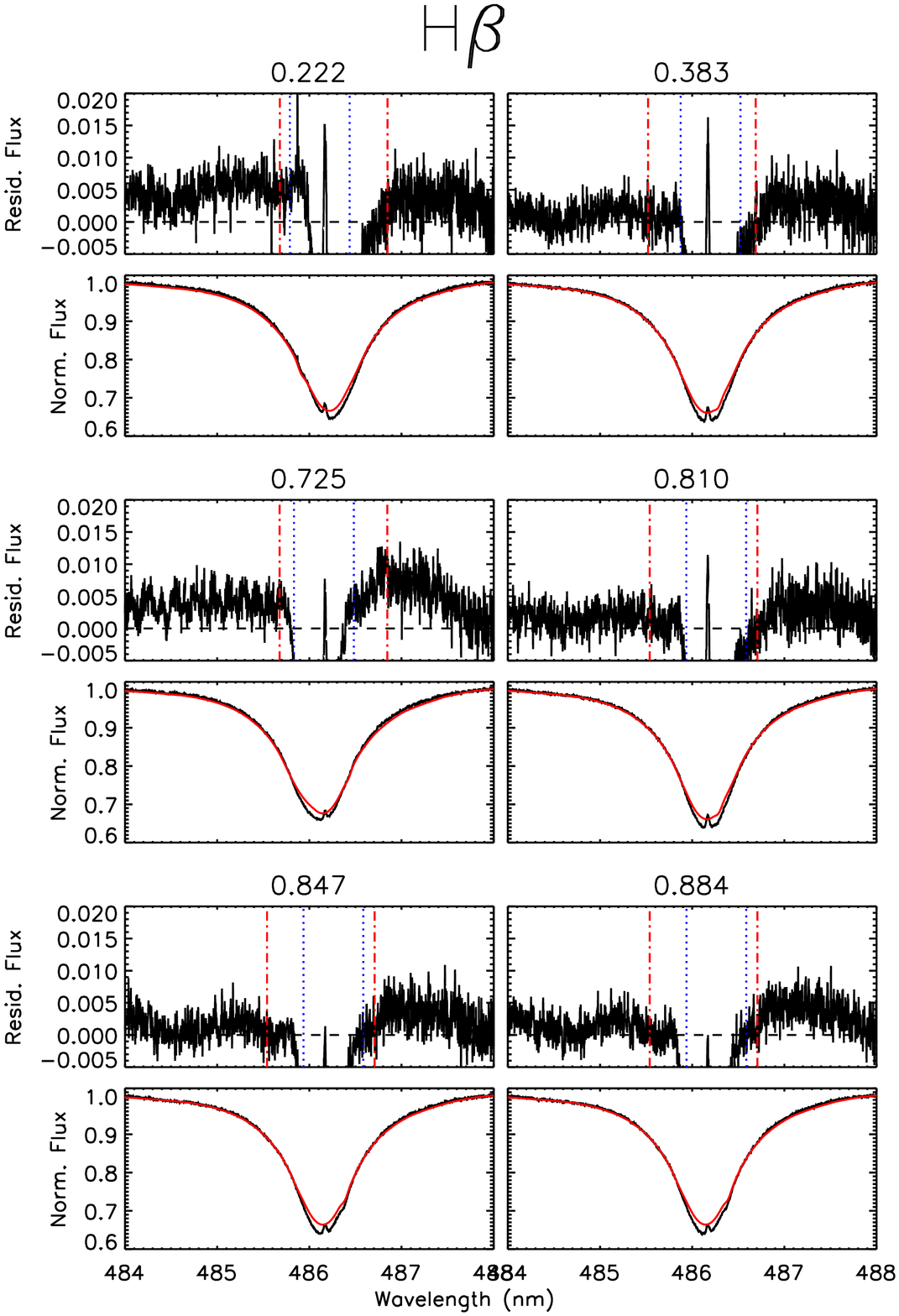} \\
\end{tabular}
      \caption[]{{\em Left panels}: Model fits to H$\alpha$ (bottom panels) and residual flux (top panels). In the bottom panels, observed flux is shown in black, synthetic line profiles in red. In the top panels, the vertical blue dotted lines show $\pm$\vsini~of NU Ori Aa, adjusted to the star's RV; red dot-dashed lines show $\pm$\rk, adjusted to NU Ori C's RV; the horizontal dashed line indicates the continuum. Note the slight flux excess in the profile wings. Plot titles indicate rotational phases. {\em Right panels}: as left, for H$\beta$. Note the absence of excess flux at high velocity.}
         \label{nuori_halpha}
   \end{figure*}

To model the surface magnetic field of the C component, we utilized the sinusoidal fit to the \bz~curve and the fit uncertainties (see Fig.\ \ref{HD37061_B_bz}) to solve Preston's equations for a centred, tilted dipole \citep{preston1967}. The rotational inclination $i_{\rm rot} = 38 \pm 5^\circ$ was obtained from the $R_*$, \vsini, and $P_{\rm rot}$ as $\sin{i} = v\sin{i} / v_{\rm eq} = v\sin{i} P_{\rm rot} /(2\pi R_*)$. The rotational inclination is clearly different from the interferometric orbital inclination. As noted above, spin-orbit misalignment is not surprising given the youth of the system and its wide, eccentric orbit.  

The obliquity $\beta = 62 \pm 6^\circ$ was determined from $i$ and the sinusoidal fitting parameters from Eqn.\ \ref{bz_fit} using the Preston $r$ parameter \citep{preston1967}:

\begin{equation}\label{r}
r = \frac{|B_0| - B_1}{|B_0| + B_1},
\end{equation}

\noindent with $\beta$ then given by \citep{preston1967}: 

\begin{equation}\label{obl}
\beta = \tan^{-1}{\left (\frac{1 - r}{1 + r}\frac{1}{\tan{(i_{\rm rot})}} \right )}.
\end{equation}

The surface polar strength of the magnetic dipole $B_{\rm d} = 7.9 \pm 1.5$~kG was calculated using \citep{preston1967}:

\begin{equation}\label{bd}
B_{\rm d} = |\langle B_z \rangle|_{\rm max}\frac{20(3 - \epsilon)}{15 + \epsilon} \frac{1}{(\cos{\beta}\cos{i_{\rm rot}} + \sin{\beta}\sin{i_{\rm rot}})},
\end{equation}

\noindent with $|\langle B_z \rangle|_{\rm max} = |B_0| + |B_1|$ and the linear limb darkening parameter $\epsilon = 0.38 \pm 0.02$ obtained from \cite{diazcordoves1995} for the \teff~and surface gravity inferred for NU Ori C from the HRD. 

\cite{2017MNRAS.471.2286S} noted that the stellar and magnetic parameters of NU Ori were very similar to those of $\xi^1$ CMa, making the failure to detect magnetospheric emission of a comparable strength in this star something of a mystery. This is partly resolved by the fact that it is NU Ori C, rather than NU Ori Aa, that is magnetic; thus, its stellar parameters are in fact {\em not} similar to those of $\xi^1$ CMa. However, the rapid rotation and strong magnetic field of NU Ori C still suggest that it may possess the ingredients necessary for a magnetosphere. The star's Kepler corotation radius \rk~\citep[Eqn. 12;][]{town2005c} is $2.6 \pm 0.1~R_*$. To determine the system's Alfv\'en radius \ra, or the furthest extent of closed magnetic loops within the magnetosphere, we used the \cite{vink2001} mass-loss rate and wind terminal velocity inferred from the star's physical parameters, obtaining $\log{({\dot{M}}/M_\odot~{\rm yr}^{-1})} = -9.3^{+0.08}_{-0.4}$ and $v_\infty = 1200^{+1300}_{-100}$~\kms (where the asymmetric error bars reflect the bistability jump at the star's \teff). We then obtain \ra~$=18^{+5}_{-1}~R_*$ \citep[Eqn. 7;][]{ud2008}, thus yielding a logarithmic ratio of $\log{(R_{\rm A}/R_{\rm K})} = 0.84^{+0.09}_{-0.03}$. 

These magnetospheric parameters are within the Centrifugal Magnetosphere (CM) regime identified by \cite{petit2013} as correlated to the presence of H$\alpha$ emission in many other magnetic early B-type stars. Inside a CM, rotational support of rigidly corotating, magnetically confined plasma prevents gravitational infall, enabling it to build up to sufficiently high densities to be detectable in the H$\alpha$ line. The emission signature of a CM is quite distinctive: typically, there are two emission bumps at high velocities (3 or 4 times \vsini), with no emission inside this range \citep[e.g.][]{bohl2011,grun2012,oks2012,rivi2013,2015MNRAS.451.1928S}. These arise due to the inability of plasma to accumulate below \rk, where the centrifugal force is weaker than gravity \citep{town2005c}. 

Emission strengths of CM stars are typically weak, around 10\% of the continuum or less. Given the flux contrast between NU Ori C and NU Ori Aa, we would then expect emission to be present at about the 1\% level. To see if such emission can be detected, synthetic line profiles were calculated using NLTE {\sc tlusty} spectra from the BSTAR2006 library \citep{2007ApJS..169...83L}, interpolated to the inferred stellar parameters, convolved with the rotational velocities, shifted to the RVs, and combined using the inferred radii of the three components. Since the emission should be much stronger in H$\alpha$ than in H$\beta$, synthetic profiles were created for both of these lines. Comparisons are shown in Fig.\ \ref{nuori_halpha}, where the 6 spectra with minimal contamination by telluric features were selected. 

The fit within $\pm$\vsini~of the Aa component is only approximate, likely due to factors unaccounted for in the model. The fit in the vicinity of the C~{\sc ii} lines near 658 nm is poor, which likely explains the apparent emission excess in the red wing of H$\alpha$. The lack of variability in the red wing further suggests that the flux excess is spurious. However, in the blue wing there is a small amount of excess flux, on the order of 1\% of the continuum, which appears in some but not in all observations. According to the VALD line lists used to extract LSD profiles, there are no spectral lines in this region. This flux excess is located outside NU Ori C's Kepler radius, as expected for magnetospheric emission. However, the flux excess extends to $\sim 10~R_{\rm C}$, which is significantly further than the $\sim 6~R_*$ that is usually seen \citep[e.g.][]{oks2012,rivi2013,grun2012}. In contrast to H$\alpha$, the wings of H$\beta$ are essentially flat, again as expected for magnetospheric emission. 

CM variability generally shows a characteristic rotationally coherent pattern of variability. Unfortunately, of the 6 observations of sufficient quality, 4 were obtained at similar phases (see plot titles in Fig.\ \ref{nuori_halpha}), making it impossible to say whether there is coherent variability following the expected pattern. Since only one magnetic pole is unambiguously seen in the \bz~curve (Fig.\ \ref{HD37061_B_bz}), the emission strength of H$\alpha$ should show only a single maximum, which should coincide with maximum \bz~at phase 0; at phase 0.5, emission strength should be at a minimum. If the flux excess in the blue wing is real, the observations acquired near phases 0.2 and 0.8 should be the strongest, while those acquired near phases 0.4 and 0.7 should be the weakest. Instead, there is no flux excess at phase 0.2, and the flux excess at 0.4 is similar to that at 0.8. This discrepancy suggests that the blue flux excess is likely also spurious. 

X-rays provide a reliable magnetospheric diagnostic, as magnetic B-type stars are typically overluminous in X-rays and generally exhibit harder X-ray spectra than non-magnetic stars \citep{2014ApJS..215...10N}. NU Ori is an apparent exception to this rule, as its X-ray spectra are actually quite soft in comparison to other magnetic early-type stars \citep{2005ApJS..160..557S,2014ApJS..215...10N}. \cite{2014ApJS..215...10N} determined the star's X-ray luminosity, corrected for absorption by the interstellar medium, via 4-temperature fits to the available {\em XMM-Newton} and {\em Chandra} data, finding $\log{L_{\rm X}} = 30.7 \pm 0.09~{\rm erg~s^{-1}}$. This agreed well with the prediction from the XADM model \citep{ud2014}, which they calculated under the assumption that the Aa component was magnetic, i.e.\ that the star was hotter, more luminous, and less strongly magnetized than is in fact the case. 

Using the stellar and magnetic parameters for the C component, XADM predicts an X-ray luminosity of $\log{L_{\rm X}} = 29.9$, i.e. the star is overluminous by about 0.8 dex. This is similar to what has been observed for other rapidly rotating, strongly magnetized early B-type stars, e.g. $\sigma$ Ori E, HR 5907, HR 7355, and HR 2949 \citep{2014ApJS..215...10N,2017IAUS..329..369F}, all of which are overluminous by 1-2 dex. This may reflect enhanced X-ray production due to centrifugal acceleration of the plasma confined in the outermost magnetosphere \citep{town2007}. Whether NU Ori's soft X-ray spectrum is anomalous in the context of X-ray overluminosity is inconclusive: while HR 5907 and HR 7355 both possess extremely hard X-ray spectra, $\sigma$ Ori E's is relatively soft \citep{2014ApJS..215...10N}. However, it should be emphasized that conclusions regarding luminosity and hardness are preliminary, as the X-ray spectra should first be corrected for the likely considerable contribution from the wind of the non-magnetic primary. 

\subsection{Magnetic fields of NU Ori Aa and Ab}\label{subsec:bfield_Aab}

   \begin{figure}
   \centering
   \includegraphics[width=8.5cm]{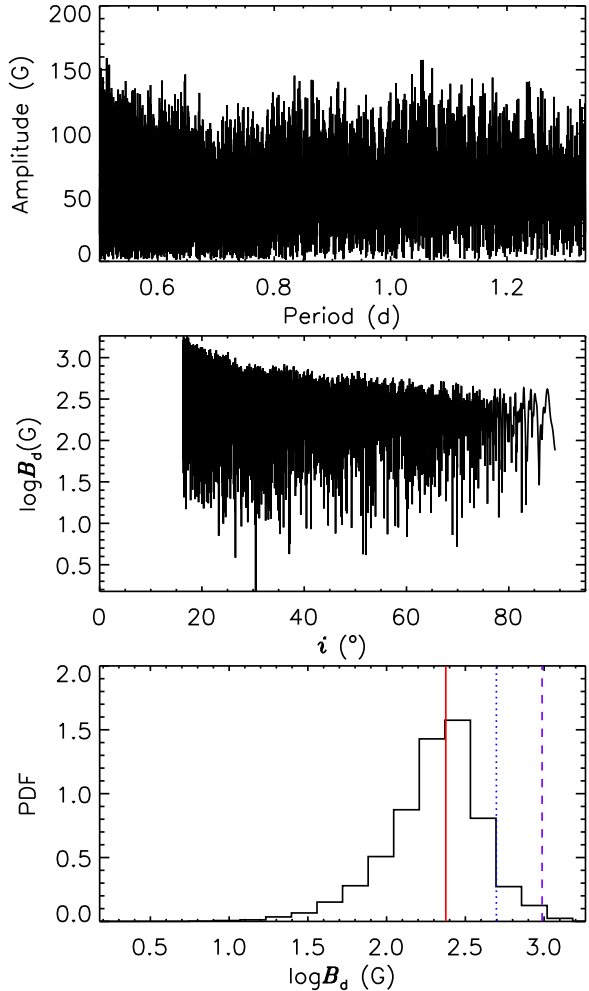}
      \caption[]{Illustration of the use of the \bz~periodogram (for the measurements from the 30 kK LSD profiles) to obtain upper limits on $B_{\rm d}$ for NU Ori Aa. {\em Top}: the \bz~periodogram, limited to the minimum and maximum periods from the star's rotational properties. The star's high \vsini~indicates that $i$ cannot be smaller than about 15$^\circ$. {\em Middle}: $B_{\rm d}$ as a function of $i$, from converting the \bz~periodogram into $i$ and $B_{\rm d}$ using the star's rotational properties and Preston's relations. {\em Bottom}: PDF of $B_{\rm d}$ after marginalizing over $i$. The solid red, dotted blue, and dashed purple lines indicate 1, 2, and 3$\sigma$ upper limits.}
         \label{Aa_bdlim}
   \end{figure}

   \begin{figure}
   \centering
   \includegraphics[width=8.5cm]{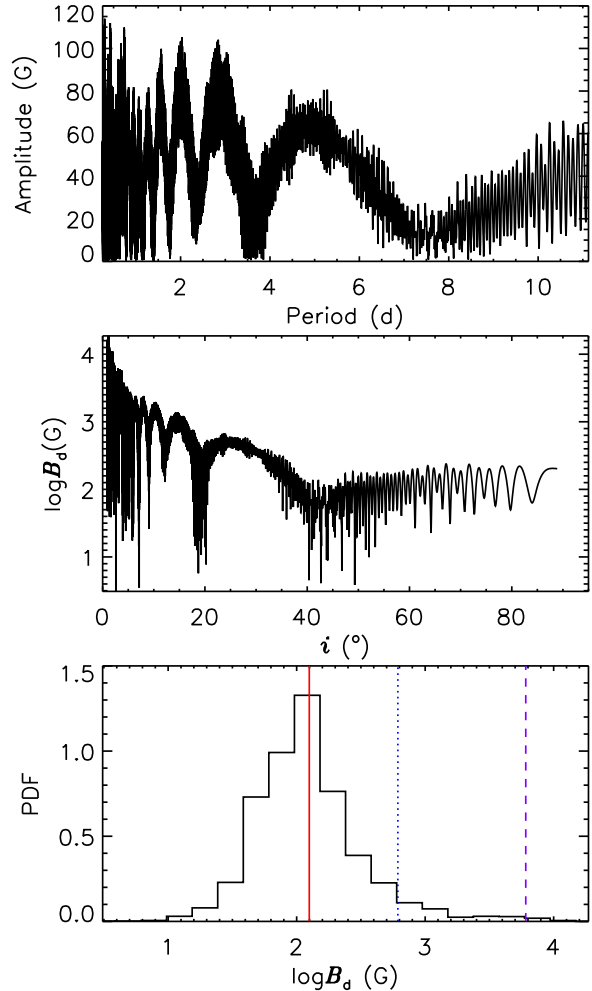}
      \caption[]{As Fig.\ \ref{Aa_bdlim}, for NU Ori Ab. The star's low \vsini~means that $i$ can be very small, leading to potentially very high values of $B_{\rm d}$.}
         \label{Ab_bdlim}
   \end{figure}

While no magnetic field is detected in either Aa or Ab, the magnetic measurements obtained here can be used to establish upper limits on their surface magnetic field strengths. This would usually be accomplished by means of direct modeling of their Stokes $V$ profiles, e.g. using a Bayesian inference approach \citep[][]{petit2012a}. However, given the likelihood that NU Ori C's contribution to Stokes $V$ is still affecting the LSD profiles of Aa and Ab, this method cannot be utilized. 

Instead, we adopt a novel method of establishing upper limits on $B_{\rm d}$, using \bz~periodograms. The assumption is that there is no signal in either the Aa or Ab time series, and that the amplitudes of their respective periodgrams thus provide a reasonable approximation of the maximum \bz~amplitude that can be hidden in the noise at any given period. Since in both cases \vsini~and $R_*$ are known, the rotational inclination can be determined directly from the period. The periodogram is automatically limited at the upper end by taking the equatorial velocity to be the same as \vsini, and at the lower end by the breakup velocity. In calculating the breakup velocity (and the inclination) we accounted for the rotational oblateness using the usual relationship for a rotating self-gravitating body \citep[e.g.][]{1928asco.book.....J}. The obliquity is then determined from Eqns.\ \ref{r} and \ref{obl}, with $B_1$ as the periodogram amplitude and $B_0$ fixed to 1 G (strictly speaking, for a pure noise \bz~time series the mean value of \bz~should be zero; however, if this is used then $\beta = 90^\circ$ for all $i$ and the parameter space is poorly explored). $B_{\rm d}$ is then determined as a function of $i$ from Eqn. \ref{bd}, with $|\langle B_z \rangle|_{\rm max} = |B_0| + |B_1|$. Overall upper limits on $B_{\rm d}$ are then determined by marginalizing the resulting $B_{\rm d}$ distribution over $i$, under the assumption that $P(i) = 0.5\sin{i}$ \citep[as is expected and generally observed, e.g.][]{2001AJ....122.2008A, 2010MNRAS.402.1380J}. 

The results of this method are shown in Figs.\ \ref{Aa_bdlim} and \ref{Ab_bdlim} for Aa and Ab, respectively. While the amplitude is relatively constant across the period ranges in question, the nature of Preston's relation leads to the inferred $B_{\rm d}$ values blowing up close to $i = 0$ or 90$^\circ$. For Aa this is mitigated by the star's high \vsini, which means that $i$ cannot be smaller than about 15$^\circ$. After marginalizing the distribution over $i$, the 1, 2, and 3$\sigma$ upper limits for Aa are respectively 230 G, 490 G, and 950 G. The corresponding upper limits for Ab are 100 G, 600 G, and 6000 G, with the more extended low-probability tail arising due to the star's effectively unconstrained inclination. 

\section{Conclusions}\label{sec:conclusions}

The line profile variability of NU Ori, seen especially in He~{\sc i} lines, can be explained by the contribution of a third star, which is also clearly detected in all interferometric observation; thus, NU Ori is an SB3 rather than an SB2. Measurement of the radial velocities of the three components via line profile fitting strongly suggests that they form a hierarchical triple: an inner system (Aab) composed of a 15 \msun~primary and a 4~\msun~secondary with an orbit of approximately 14 d, which is in turn orbited by an 8 \msun~tertiary (C) with an orbital period of about 476 d. Orbital modelling of RVs and interferometry indicates that the inner system is approximatly circular, while the outer orbit is mildly eccentric. The orbital axes appear to be approximately aligned: the inclination of the orbital axis of AC, as determined via interferometry, is $70.1 \pm 0.9^\circ$, and the projected mass function for the inner binary yields $70 \pm 4^\circ$ based on the masses inferred from the HRD.

In principle, highly precise masses for the three stars can be obtained via interferometry. Unfortunately, the distance is not known with sufficient precision. Future Gaia data releases that account for multiplicity should yield a more accurate parallax for this system, with which a more precise comparison can be made between the astrometric masses and the masses inferred from evolutionary models. These data will also enable the orbital properties of the system to be further constrained. 

We have found that the previously reported magnetic field, which had been attributed to the  B0.5~V primary, is in fact hosted by the B2~V C component. This is supported by the confinement of the Zeeman signature within the line profile of the C component, and the movement of the Zeeman signature with the RV of the C rather than the Aa component. This has motivated a re-analysis of the rotational and magnetic properties of NU Ori. The previously reported rotational period of  $\sim1.1$~d is confirmed, however, the inferred surface magnetic dipole strength is much stronger, about 8 kG. The rotational axis of the C component appears to be misaligned by about 30$^\circ$ relative to the AC orbital axis. 

NU Ori C's rapid rotation and strong magnetic confinement are consistent with parameters typically associated with H$\alpha$ emission, which would be expected at about 1\% of the continuum level of the SB3 system. We find evidence of emission at the expected magnitude via comparison of synthetic line profiles to H$\alpha$ and H$\beta$ observations, although this will need to be confirmed. Re-evaluation of the star's expected X-ray luminosity shows that the system may show a degree of overluminosity similar to that of other rapidly rotating, strongly magnetic $\sigma$ Ori E-type stars. This preliminary conclusion should be revisited by correcting the X-ray spectra for the influence of the non-magnetic primary, and isolating the X-ray spectrum of NU Ori C. 

The 3$\sigma$ upper limits on the surface dipole magnetic fields of the Aa and Ab components are about 1 and 6 kG, respectively (i.e., both stars are less, and probably much less, magnetic than NU Ori C), where we utilized a novel analytic method that combines rotational information to convert period spectra obtained from \bz~time series into probability density functions of $B_{\rm d}$ marginalized over $i$. 

\section*{Acknowledgements}
This work has made use of the VALD database, operated at Uppsala University, the Institute of Astronomy RAS in Moscow, and the University of Vienna. This work is based on observations obtained at the Canada-France-Hawaii Telescope (CFHT) which is operated by the National Research Council of Canada, the Institut National des Sciences de l'Univers of the Centre National de la Recherche Scientifique of France, and the University of Hawaii. This work is based on observations collected at the European Organisation for Astronomical Research in the Southern Hemisphere under ESO programmes 60.A-0209(A), 092.C-0542(A), 094.C-0175(A), 094.C-0397(A), 096.D-0518(A), and 0100.C-0597(A). This work has made use of data from the European Space Agency (ESA) mission {\em Gaia} (\url{https://www.cosmos.esa.int/gaia}), processed by the {\em Gaia} Data Processing and Analysis Consortium (DPAC, \url{https://www.cosmos.esa.int/web/gaia/dpac/consortium}). Funding for the DPAC has been provided by national institutions, in particular the institutions participating in the {\it Gaia} Multilateral Agreement. This work utilized data obtained with GRAVITY. GRAVITY is developed in collaboration by the Max Planck Institute for Extraterrestrial Physics, LESIA of Paris Observatory/CNRS/UPMC/University of Paris Diderot and IPAG Universit\'e Grenoble, the Max Planck Institute for Astronomy, the University of Cologne, and the Centro Multidisciplinar de Astrof\'isica Lisbon and Porto. This project used the facilities of SIMBAD. All authors acknowledge the advice and assistance provided on this and related projects by the members of the BinaMIcS and MiMeS collaborations. MS acknowledges the support of the Annie Jump Cannon Fellowship endowed by the Mount Cuba Astronomical Observatory, and of the Natural Sciences and Engineering Research Council (NSERC) Postdoctoral Fellowship program. GAW acknowledges support from an NSERC Discovery Grant. OK acknowledges financial support from the Knut and Alice Wallenberg Foundation, the Swedish Research Council, and the Swedish National Space Board. PG acknowledges support from FCT-Portugal with reference UID-FIS-00099-2013.

\bibliography{bib_dat.bib}{}

\begin{thebibliography}{69}
\expandafter\ifx\csname natexlab\endcsname\relax\def\natexlab#1{#1}\fi

\bibitem[{{Abt}(2001)}]{2001AJ....122.2008A}
{Abt} H.~A., 2001, \aj, 122, 2008

\bibitem[{{Abt} {et~al}\mbox{.}(1991){Abt}, {Wang}, \&
  {Cardona}}]{1991ApJ...367..155A}
{Abt} H.~A., {Wang} R., {Cardona} O., 1991, \apj, 367, 155

\bibitem[{{Alecian} {et~al}\mbox{.}(2015){Alecian}, {Neiner}, {Wade}, {Mathis},
  {Bohlender}, {C{\'e}bron}, {Folsom}, {Grunhut}, {Le Bouquin}, {Petit},
  {Sana}, {Tkachenko}, \& {ud-Doula}}]{2015IAUS..307..330A}
{Alecian} E. {et~al.}, 2015, in IAU Symposium, Vol. 307, New Windows on Massive
  Stars, pp. 330--335

\bibitem[{{Babel} \& {Montmerle}(1997)}]{bm1997}
{Babel} J., {Montmerle} T., 1997, \apjl, 485, L29

\bibitem[{{Bohlender} \& {Monin}(2011)}]{bohl2011}
{Bohlender} D.~A., {Monin} D., 2011, \aj, 141, 169

\bibitem[{{Bonnell} \& {Bate}(1994)}]{1994MNRAS.271..999B}
{Bonnell} I.~A., {Bate} M.~R., 1994, \mnras, 271, 999

\bibitem[{{Commer{\c c}on} {et~al}\mbox{.}(2011){Commer{\c c}on}, {Hennebelle},
  \& {Henning}}]{2011ApJ...742L...9C}
{Commer{\c c}on} B., {Hennebelle} P., {Henning} T., 2011, \apjl, 742, L9

\bibitem[{{D\'iaz-Cordov\'es} {et~al}\mbox{.}(1995){D\'iaz-Cordov\'es},
  {Claret}, \& {Gim\'enez}}]{diazcordoves1995}
{D\'iaz-Cordov\'es} J., {Claret} A., {Gim\'enez} A., 1995, \aaps, 110, 329

\bibitem[{{Donati} {et~al}\mbox{.}(2006){Donati}, {Howarth}, {Jardine},
  {Petit}, {Catala}, {Landstreet}, {Bouret}, {Alecian}, {Barnes}, {Forveille},
  {Paletou}, \& {Manset}}]{2006MNRAS.370..629D}
{Donati} J.-F. {et~al.}, 2006, \mnras, 370, 629

\bibitem[{{Donati} {et~al}\mbox{.}(1997){Donati}, {Semel}, {Carter}, {Rees}, \&
  {Collier Cameron}}]{d1997}
{Donati} J.-F., {Semel} M., {Carter} B.~D., {Rees} D.~E., {Collier Cameron} A.,
  1997, MNRAS, 291, 658

\bibitem[{{Donati} {et~al}\mbox{.}(1992){Donati}, {Semel}, \&
  {Rees}}]{1992AA...265..669D}
{Donati} J.-F., {Semel} M., {Rees} D.~E., 1992, \aap, 265, 669

\bibitem[{{Ekstr{\"o}m} {et~al}\mbox{.}(2012){Ekstr{\"o}m}, {Georgy},
  {Eggenberger}, {Meynet}, {Mowlavi}, {Wyttenbach}, {Granada}, {Decressin},
  {Hirschi}, {Frischknecht}, {Charbonnel}, \& {Maeder}}]{ekstrom2012}
{Ekstr{\"o}m} S. {et~al.}, 2012, \aap, 537, A146

\bibitem[{{Fletcher} {et~al}\mbox{.}(2017){Fletcher}, {Petit}, {Naz{\'e}},
  {Wade}, {Townsend}, {Owocki}, {Cohen}, {David-Uraz}, \&
  {Shultz}}]{2017IAUS..329..369F}
{Fletcher} C.~L. {et~al.}, 2017, in IAU Symposium, Vol. 329, The Lives and
  Death-Throes of Massive Stars, {Eldridge} J.~J., {Bray} J.~C., {McClelland}
  L.~A.~S., {Xiao} L., eds., pp. 369--372

\bibitem[{{Fukui} {et~al}\mbox{.}(2018){Fukui}, {Torii}, {Hattori},
  {Nishimura}, {Ohama}, {Shimajiri}, {Shima}, {Habe}, {Sano}, {Kohno},
  {Yamamoto}, {Tachihara}, \& {Onishi}}]{2018ApJ...859..166F}
{Fukui} Y. {et~al.}, 2018, \apj, 859, 166

\bibitem[{{Gaia Collaboration} {et~al}\mbox{.}(2018){Gaia Collaboration},
  {Brown}, {Vallenari}, {Prusti}, {de Bruijne}, {Babusiaux}, {Bailer-Jones},
  {Biermann}, {Evans}, {Eyer}, \& et~al.}]{2018A&A...616A...1G}
{Gaia Collaboration} {et~al.}, 2018, \aap, 616, A1

\bibitem[{{Gaia Collaboration} {et~al}\mbox{.}(2016){Gaia Collaboration},
  {Brown}, {Vallenari}, {Prusti}, {de Bruijne}, {Mignard}, {Drimmel},
  {Babusiaux}, {Bailer-Jones}, {Bastian}, \& et~al.}]{2016A&A...595A...2G}
{Gaia Collaboration} {et~al.}, 2016, \aap, 595, A2

\bibitem[{{Gelman} \& {Rubin}(1992)}]{1992StaSc...7..457G}
{Gelman} A., {Rubin} D.~B., 1992, Statistical Science, 7, 457

\bibitem[{{Gonz{\'a}lez} \& {Levato}(2006)}]{2006AA...448..283G}
{Gonz{\'a}lez} J.~F., {Levato} H., 2006, \aap, 448, 283

\bibitem[{{Gravity Collaboration} {et~al}\mbox{.}(2017){Gravity Collaboration},
  {Abuter}, {Accardo}, {Amorim}, {Anugu}, {{\'A}vila}, {Azouaoui}, {Benisty},
  {Berger}, {Blind}, {Bonnet}, {Bourget}, {Brandner}, {Brast}, {Buron},
  {Burtscher}, {Cassaing}, {Chapron}, {Choquet}, {Cl{\'e}net}, {Collin},
  {Coud{\'e} Du Foresto}, {de Wit}, {de Zeeuw}, {Deen},
  {Delplancke-Str{\"o}bele}, {Dembet}, {Derie}, {Dexter}, {Duvert}, {Ebert},
  {Eckart}, {Eisenhauer}, {Esselborn}, {F{\'e}dou}, {Finger}, {Garcia}, {Garcia
  Dabo}, {Garcia Lopez}, {Gendron}, {Genzel}, {Gillessen}, {Gonte}, {Gordo},
  {Grould}, {Gr{\"o}zinger}, {Guieu}, {Haguenauer}, {Hans}, {Haubois}, {Haug},
  {Haussmann}, {Henning}, {Hippler}, {Horrobin}, {Huber}, {Hubert}, {Hubin},
  {Hummel}, {Jakob}, {Janssen}, {Jochum}, {Jocou}, {Kaufer}, {Kellner},
  {Kendrew}, {Kern}, {Kervella}, {Kiekebusch}, {Klein}, {Kok}, {Kolb}, {Kulas},
  {Lacour}, {Lapeyr{\`e}re}, {Lazareff}, {Le Bouquin}, {L{\`e}na}, {Lenzen},
  {L{\'e}v{\^e}que}, {Lippa}, {Magnard}, {Mehrgan}, {Mellein}, {M{\'e}rand},
  {Moreno-Ventas}, {Moulin}, {M{\"u}ller}, {M{\"u}ller}, {Neumann}, {Oberti},
  {Ott}, {Pallanca}, {Panduro}, {Pasquini}, {Paumard}, {Percheron}, {Perraut},
  {Perrin}, {Pfl{\"u}ger}, {Pfuhl}, {Phan Duc}, {Plewa}, {Popovic}, {Rabien},
  {Ram{\'{\i}}rez}, {Ramos}, {Rau}, {Riquelme}, {Rohloff}, {Rousset},
  {Sanchez-Bermudez}, {Scheithauer}, {Sch{\"o}ller}, {Schuhler}, {Spyromilio},
  {Straubmeier}, {Sturm}, {Suarez}, {Tristram}, {Ventura}, {Vincent},
  {Waisberg}, {Wank}, {Weber}, {Wieprecht}, {Wiest}, {Wiezorrek}, {Wittkowski},
  {Woillez}, {Wolff}, {Yazici}, {Ziegler}, \& {Zins}}]{2017A&A...602A..94G}
{Gravity Collaboration} {et~al.}, 2017, \aap, 602, A94

\bibitem[{{Gravity Collaboration} {et~al}\mbox{.}(2018){Gravity Collaboration},
  {Pfuhl}, \& {Karl}}]{gravity2018}
{Gravity Collaboration}, {Pfuhl} O., {Karl} M., 2018, submitted

\bibitem[{{Grellmann} {et~al}\mbox{.}(2013){Grellmann}, {Preibisch}, {Ratzka},
  {Kraus}, {Helminiak}, \& {Zinnecker}}]{2013A&A...550A..82G}
{Grellmann} R., {Preibisch} T., {Ratzka} T., {Kraus} S., {Helminiak} K.~G.,
  {Zinnecker} H., 2013, \aap, 550, A82

\bibitem[{{Grunhut} {et~al}\mbox{.}(2012){Grunhut}, {Rivinius}, {Wade},
  {Townsend}, {Marcolino}, {Bohlender}, {Szeifert}, {Petit}, {Matthews},
  {Rowe}, {Moffat}, {Kallinger}, {Kuschnig}, {Guenther}, {Rucinski},
  {Sasselov}, \& {Weiss}}]{grun2012}
{Grunhut} J.~H. {et~al.}, 2012, \mnras, 419, 1610

\bibitem[{{Grunhut} {et~al}\mbox{.}(2017){Grunhut}, {Wade}, {Neiner}, {Oksala},
  {Petit}, {Alecian}, {Bohlender}, {Bouret}, {Henrichs}, {Hussain},
  {Kochukhov}, \& {MiMeS Collaboration}}]{2017MNRAS.465.2432G}
{Grunhut} J.~H. {et~al.}, 2017, \mnras, 465, 2432

\bibitem[{{Hut}(1981)}]{1981AA....99..126H}
{Hut} P., 1981, \aap, 99, 126

\bibitem[{{Jackson} \& {Jeffries}(2010)}]{2010MNRAS.402.1380J}
{Jackson} R.~J., {Jeffries} R.~D., 2010, \mnras, 402, 1380

\bibitem[{{Jeans}(1928)}]{1928asco.book.....J}
{Jeans} J.~H., 1928, {Astronomy and cosmogony}. Cambridge [Eng.] The University
  press

\bibitem[{{K{\"o}hler} {et~al}\mbox{.}(2006){K{\"o}hler}, {Petr-Gotzens},
  {McCaughrean}, {Bouvier}, {Duch{\^e}ne}, {Quirrenbach}, \&
  {Zinnecker}}]{2006A&A...458..461K}
{K{\"o}hler} R., {Petr-Gotzens} M.~G., {McCaughrean} M.~J., {Bouvier} J.,
  {Duch{\^e}ne} G., {Quirrenbach} A., {Zinnecker} H., 2006, \aap, 458, 461

\bibitem[{{Kounkel}(2017)}]{2017PhDT........55K}
{Kounkel} M., 2017, PhD thesis, University of Michigan

\bibitem[{{Kupka} {et~al}\mbox{.}(1999){Kupka}, {Piskunov}, {Ryabchikova},
  {Stempels}, \& {Weiss}}]{kupka1999}
{Kupka} F.~G., {Piskunov} N., {Ryabchikova} T.~A., {Stempels} H.~C., {Weiss}
  W.~W., 1999, \aaps, 138, 119

\bibitem[{{Kupka} {et~al}\mbox{.}(2000){Kupka}, {Ryabchikova}, {Piskunov},
  {Stempels}, \& {Weiss}}]{kupka2000}
{Kupka} F.~G., {Ryabchikova} T.~A., {Piskunov} N.~E., {Stempels} H.~C., {Weiss}
  W.~W., 2000, Baltic Astronomy, 9, 590

\bibitem[{{Lanz} \& {Hubeny}(2007)}]{2007ApJS..169...83L}
{Lanz} T., {Hubeny} I., 2007, \apjs, 169, 83

\bibitem[{{Le Bouquin} {et~al}\mbox{.}(2011){Le Bouquin}, {Berger}, {Lazareff},
  {Zins}, {Haguenauer}, {Jocou}, {Kern}, {Millan-Gabet}, {Traub}, {Absil},
  {Augereau}, {Benisty}, {Blind}, {Bonfils}, {Bourget}, {Delboulbe},
  {Feautrier}, {Germain}, {Gitton}, {Gillier}, {Kiekebusch}, {Kluska},
  {Knudstrup}, {Labeye}, {Lizon}, {Monin}, {Magnard}, {Malbet}, {Maurel},
  {M{\'e}nard}, {Micallef}, {Michaud}, {Montagnier}, {Morel}, {Moulin},
  {Perraut}, {Popovic}, {Rabou}, {Rochat}, {Rojas}, {Roussel}, {Roux},
  {Stadler}, {Stefl}, {Tatulli}, \& {Ventura}}]{2011A&A...535A..67L}
{Le Bouquin} J.-B. {et~al.}, 2011, \aap, 535, A67

\bibitem[{{Le Bouquin} {et~al}\mbox{.}(2017){Le Bouquin}, {Sana}, {Gosset}, {De
  Becker}, {Duvert}, {Absil}, {Anthonioz}, {Berger}, {Ertel}, {Grellmann},
  {Guieu}, {Kervella}, {Rabus}, \& {Willson}}]{2017A&A...601A..34L}
{Le Bouquin} J.-B. {et~al.}, 2017, \aap, 601, A34

\bibitem[{{Lenz} \& {Breger}(2005)}]{2005CoAst.146...53L}
{Lenz} P., {Breger} M., 2005, Communications in Asteroseismology, 146, 53

\bibitem[{{Mathys}(1989)}]{mat1989}
{Mathys} G., 1989, FCPh, 13, 143

\bibitem[{{Mayne} \& {Naylor}(2008)}]{2008MNRAS.386..261M}
{Mayne} N.~J., {Naylor} T., 2008, \mnras, 386, 261

\bibitem[{{Mazeh} \& {Shaham}(1979)}]{1979A&A....77..145M}
{Mazeh} T., {Shaham} J., 1979, \aap, 77, 145

\bibitem[{{M{\'e}rand} {et~al}\mbox{.}(2014){M{\'e}rand}, {Abuter},
  {Aller-Carpentier}, {Andolfato}, {Alonso}, {Berger}, {Blanchard}, {Boffin},
  {Bourget}, {Bristow}, {Cid}, {de Wit}, {del Valle},
  {Delplancke-Str{\"o}bele}, {Derie}, {Faundez}, {Ertel}, {Grellmann},
  {Gitton}, {Glindemann}, {Guajardo}, {Guieu}, {Guisard}, {Guniat},
  {Haguenauer}, {Herrera}, {Hummel}, {La Fuente}, {Lopez}, {Mardones}, {Morel},
  {M{\"u}ller}, {Percheron}, {Duc}, {Pino}, {Poupar}, {Pozna}, {Ramirez},
  {Rengaswamy}, {Rivas}, {Rivinius}, {Segovia}, {Schmid}, {Sch{\"o}ller},
  {Schuhler}, {Woillez}, \& {Wittkowski}}]{2014SPIE.9146E..0JM}
{M{\'e}rand} A. {et~al.}, 2014, in \procspie, Vol. 9146, Optical and Infrared
  Interferometry IV, p. 91460J

\bibitem[{{Morrell} \& {Levato}(1991)}]{1991ApJS...75..965M}
{Morrell} N., {Levato} H., 1991, \apjs, 75, 965

\bibitem[{{Naz{\'e}} {et~al}\mbox{.}(2014){Naz{\'e}}, {Petit}, {Rinbrand},
  {Cohen}, {Owocki}, {ud-Doula}, \& {Wade}}]{2014ApJS..215...10N}
{Naz{\'e}} Y., {Petit} V., {Rinbrand} M., {Cohen} D., {Owocki} S., {ud-Doula}
  A., {Wade} G.~A., 2014, \apjs, 215, 10

\bibitem[{{Oksala} {et~al}\mbox{.}(2012){Oksala}, {Wade}, {Townsend}, {Owocki},
  {Kochukhov}, {Neiner}, {Alecian}, \& {Grunhut}}]{oks2012}
{Oksala} M.~E., {Wade} G.~A., {Townsend} R.~H.~D., {Owocki} S.~P., {Kochukhov}
  O., {Neiner} C., {Alecian} E., {Grunhut} J., 2012, MNRAS, 419, 959

\bibitem[{{Pablo} {et~al}\mbox{.}(in prep.){Pablo}, {Shultz}, {Fuller}, {Wade},
  {Mathis}, \& {Paunzen}}]{pablo_epslup}
{Pablo} H., {Shultz} M., {Fuller} J., {Wade} G.~A., {Mathis} S., {Paunzen} E.,
  in prep., \mnras, in prep., 1

\bibitem[{{Petit} {et~al}\mbox{.}(2013){Petit}, {Owocki}, {Wade}, {Cohen},
  {Sundqvist}, {Gagn{\'e}}, {Ma{\'{\i}}z Apell{\'a}niz}, {Oksala}, {Bohlender},
  {Rivinius}, {Henrichs}, {Alecian}, {Townsend}, {ud-Doula}, \& {MiMeS
  Collaboration}}]{petit2013}
{Petit} V. {et~al.}, 2013, \mnras, 429, 398

\bibitem[{{Petit} \& {Wade}(2012)}]{petit2012a}
{Petit} V., {Wade} G.~A., 2012, MNRAS, 420, 773

\bibitem[{{Petit} {et~al}\mbox{.}(2008){Petit}, {Wade}, {Drissen}, {Montmerle},
  \& {Alecian}}]{2008MNRAS.387L..23P}
{Petit} V., {Wade} G.~A., {Drissen} L., {Montmerle} T., {Alecian} E., 2008,
  \mnras, 387, L23

\bibitem[{{Piskunov} {et~al}\mbox{.}(1995){Piskunov}, {Kupka}, {Ryabchikova},
  {Weiss}, \& {Jeffery}}]{piskunov1995}
{Piskunov} N.~E., {Kupka} F., {Ryabchikova} T.~A., {Weiss} W.~W., {Jeffery}
  C.~S., 1995, \aaps, 112, 525

\bibitem[{{Preston}(1967)}]{preston1967}
{Preston} G.~W., 1967, \apj, 150, 547

\bibitem[{{Rivinius} {et~al}\mbox{.}(2013){Rivinius}, {Townsend}, {Kochukhov},
  {{\v S}tefl}, {Baade}, {Barrera}, \& {Szeifert}}]{rivi2013}
{Rivinius} T., {Townsend} R.~H.~D., {Kochukhov} O., {{\v S}tefl} S., {Baade}
  D., {Barrera} L., {Szeifert} T., 2013, \mnras, 429, 177

\bibitem[{{Ryabchikova} {et~al}\mbox{.}(2015){Ryabchikova}, {Piskunov},
  {Kurucz}, {Stempels}, {Heiter}, {Pakhomov}, \&
  {Barklem}}]{2015PhyS...90e4005R}
{Ryabchikova} T., {Piskunov} N., {Kurucz} R.~L., {Stempels} H.~C., {Heiter} U.,
  {Pakhomov} Y., {Barklem} P.~S., 2015, \physscr, 90, 054005

\bibitem[{{Ryabchikova} {et~al}\mbox{.}(1997){Ryabchikova}, {Piskunov},
  {Kupka}, \& {Weiss}}]{ryabchikova1997}
{Ryabchikova} T.~A., {Piskunov} N.~E., {Kupka} F., {Weiss} W.~W., 1997, Baltic
  Astronomy, 6, 244

\bibitem[{{Sandstrom} {et~al}\mbox{.}(2007){Sandstrom}, {Peek}, {Bower},
  {Bolatto}, \& {Plambeck}}]{2007ApJ...667.1161S}
{Sandstrom} K.~M., {Peek} J.~E.~G., {Bower} G.~C., {Bolatto} A.~D., {Plambeck}
  R.~L., 2007, \apj, 667, 1161

\bibitem[{{Schneider} {et~al}\mbox{.}(2016){Schneider}, {Podsiadlowski},
  {Langer}, {Castro}, \& {Fossati}}]{2016MNRAS.457.2355S}
{Schneider} F.~R.~N., {Podsiadlowski} P., {Langer} N., {Castro} N., {Fossati}
  L., 2016, \mnras, 457, 2355

\bibitem[{{Shultz} {et~al}\mbox{.}(2015){Shultz}, {Rivinius}, {Folsom}, {Wade},
  {Townsend}, {Sikora}, {Grunhut}, {Stahl}, \& {MiMeS
  Collaboration}}]{2015MNRAS.449.3945S}
{Shultz} M. {et~al.}, 2015, \mnras, 449, 3945

\bibitem[{{Shultz} {et~al}\mbox{.}(2018{\natexlab{a}}){Shultz}, {Rivinius},
  {Wade}, {Alecian}, \& {Petit}}]{2018MNRAS.475..839S}
{Shultz} M., {Rivinius} T., {Wade} G.~A., {Alecian} E., {Petit} V.,
  2018{\natexlab{a}}, \mnras, 475, 839

\bibitem[{{Shultz} {et~al}\mbox{.}(2017){Shultz}, {Wade}, {Rivinius}, {Neiner},
  {Henrichs}, {Marcolino}, \& {MiMeS Collaboration}}]{2017MNRAS.471.2286S}
{Shultz} M., {Wade} G.~A., {Rivinius} T., {Neiner} C., {Henrichs} H.,
  {Marcolino} W., {MiMeS Collaboration}, 2017, \mnras, 471, 2286

\bibitem[{{Shultz} {et~al}\mbox{.}(2018{\natexlab{b}}){Shultz}, {Wade},
  {Rivinius}, {Neiner}, {Alecian}, {Bohlender}, {Monin}, {Sikora}, {MiMeS
  Collaboration}, \& {BinaMIcS Collaboration}}]{2018MNRAS.475.5144S}
{Shultz} M.~E. {et~al.}, 2018{\natexlab{b}}, \mnras, 475, 5144

\bibitem[{{Sikora} {et~al}\mbox{.}(2015){Sikora}, {Wade}, {Bohlender},
  {Neiner}, {Oksala}, {Shultz}, {Cohen}, {ud-Doula}, {Grunhut}, {Monin},
  {Owocki}, {Petit}, {Rivinus}, \& {Townsend}}]{2015MNRAS.451.1928S}
{Sikora} J. {et~al.}, 2015, \mnras, 451, 1928

\bibitem[{{Sim{\'o}n-D{\'{\i}}az} {et~al}\mbox{.}(2011){Sim{\'o}n-D{\'{\i}}az},
  {Garc{\'{\i}}a-Rojas}, {Esteban}, {Stasi{\'n}ska}, {L{\'o}pez-S{\'a}nchez},
  \& {Morisset}}]{2011AA...530A..57S}
{Sim{\'o}n-D{\'{\i}}az} S., {Garc{\'{\i}}a-Rojas} J., {Esteban} C.,
  {Stasi{\'n}ska} G., {L{\'o}pez-S{\'a}nchez} A.~R., {Morisset} C., 2011, \aap,
  530, A57

\bibitem[{{Sim{\'o}n-D{\'{\i}}az} {et~al}\mbox{.}(2017){Sim{\'o}n-D{\'{\i}}az},
  {Godart}, {Castro}, {Herrero}, {Aerts}, {Puls}, {Telting}, \&
  {Grassitelli}}]{2017A&A...597A..22S}
{Sim{\'o}n-D{\'{\i}}az} S., {Godart} M., {Castro} N., {Herrero} A., {Aerts} C.,
  {Puls} J., {Telting} J., {Grassitelli} L., 2017, \aap, 597, A22

\bibitem[{{Sim{\'o}n-D{\'{\i}}az} \& {Herrero}(2014)}]{2014A&A...562A.135S}
{Sim{\'o}n-D{\'{\i}}az} S., {Herrero} A., 2014, \aap, 562, A135

\bibitem[{{Stelzer} {et~al}\mbox{.}(2005){Stelzer}, {Flaccomio}, {Montmerle},
  {Micela}, {Sciortino}, {Favata}, {Preibisch}, \&
  {Feigelson}}]{2005ApJS..160..557S}
{Stelzer} B., {Flaccomio} E., {Montmerle} T., {Micela} G., {Sciortino} S.,
  {Favata} F., {Preibisch} T., {Feigelson} E.~D., 2005, \apjs, 160, 557

\bibitem[{{Townsend} \& {Owocki}(2005)}]{town2005c}
{Townsend} R.~H.~D., {Owocki} S.~P., 2005, \mnras, 357, 251

\bibitem[{{Townsend} {et~al}\mbox{.}(2007){Townsend}, {Owocki}, \&
  {Ud-Doula}}]{town2007}
{Townsend} R.~H.~D., {Owocki} S.~P., {Ud-Doula} A., 2007, \mnras, 382, 139

\bibitem[{{ud-Doula} {et~al}\mbox{.}(2014){ud-Doula}, {Owocki}, {Townsend},
  {Petit}, \& {Cohen}}]{ud2014}
{ud-Doula} A., {Owocki} S., {Townsend} R., {Petit} V., {Cohen} D., 2014,
  \mnras, 441, 3600

\bibitem[{{ud-Doula} \& {Owocki}(2002)}]{ud2002}
{ud-Doula} A., {Owocki} S.~P., 2002, ApJ, 576, 413

\bibitem[{{ud-Doula} {et~al}\mbox{.}(2008){ud-Doula}, {Owocki}, \&
  {Townsend}}]{ud2008}
{ud-Doula} A., {Owocki} S.~P., {Townsend} R.~H.~D., 2008, MNRAS, 385, 97

\bibitem[{{Vink} {et~al}\mbox{.}(2001){Vink}, {de Koter}, \&
  {Lamers}}]{vink2001}
{Vink} J.~S., {de Koter} A., {Lamers} H.~J.~G.~L.~M., 2001, \aap, 369, 574

\bibitem[{{Wade} {et~al}\mbox{.}(2016){Wade}, {Neiner}, {Alecian}, {Grunhut},
  {Petit}, {Batz}, {Bohlender}, {Cohen}, {Henrichs}, {Kochukhov}, {Landstreet},
  {Manset}, {Martins}, {Mathis}, {Oksala}, {Owocki}, {Rivinius}, {Shultz},
  {Sundqvist}, {Townsend}, {ud-Doula}, {Bouret}, {Braithwaite}, {Briquet},
  {Carciofi}, {David-Uraz}, {Folsom}, {Fullerton}, {Leroy}, {Marcolino},
  {Moffat}, {Naz{\'e}}, {Louis}, {Auri{\`e}re}, {Bagnulo}, {Bailey},
  {Barb{\'a}}, {Blaz{\`e}re}, {B{\"o}hm}, {Catala}, {Donati}, {Ferrario},
  {Harrington}, {Howarth}, {Ignace}, {Kaper}, {L{\"u}ftinger}, {Prinja},
  {Vink}, {Weiss}, \& {Yakunin}}]{2016MNRAS.456....2W}
{Wade} G.~A. {et~al.}, 2016, \mnras, 456, 2

\bibitem[{{Yakunin} {et~al}\mbox{.}(2015){Yakunin}, {Wade}, {Bohlender},
  {Kochukhov}, {Marcolino}, {Shultz}, {Monin}, {Grunhut}, {Sitnova}, {Tsymbal},
  \& {MiMeS Collaboration}}]{2015MNRAS.447.1418Y}
{Yakunin} I. {et~al.}, 2015, \mnras, 447, 1418

\end{thebibliography}

\end{document}